\pdfoutput=1

\documentclass[prb,twocolumn,twoside,flushbottom,balancelastpage,superscriptaddress,showpacs,amsmath,amssymb,amsfonts]{revtex4}

\usepackage[final,notref,notcite]{showkeys} 
\usepackage{graphicx} 
\usepackage{color} 
\usepackage{amsmath} 
\usepackage{amssymb} 
\usepackage{amsfonts} 
\usepackage{mathtools} 
\usepackage[squaren]{SIunits} 
\usepackage[sort&compress]{natbib} 
\usepackage{natmove} 
\usepackage{hyperref} 
\usepackage{dcolumn} 
\usepackage{multirow} 
\usepackage{ifpdf} 
\usepackage{wasysym} 
\usepackage{hyphenat} 
\usepackage[caption=false]{subfig} 
\usepackage{caption3} 

\ifpdf
\fi

\begin{document}

\title{Impact of octahedral rotations on\texorpdfstring{\\}{ }Ruddlesden--Popper phases of antiferrodistortive perovskites}

\author{Daniel~A.~Freedman}
\affiliation{Laboratory of Atomic and Solid State Physics, Cornell University, Ithaca, NY 14853}
\affiliation{Cornell Center for Materials Research, Cornell University, Ithaca, NY 14853}
\author{T.A.~Arias}
\affiliation{Laboratory of Atomic and Solid State Physics, Cornell University, Ithaca, NY 14853}
\affiliation{Cornell Center for Materials Research, Cornell University, Ithaca, NY 14853}

\date{December 31, 2008}

\begin{abstract}
  This work presents the most detailed and extensive theoretical study
  to date of the structural configurations of Ruddlesden--Popper (RP)
  phases in antiferrodistortive (AFD) perovskites and formulates a
  program of study which can be pursued for RP phases of any AFD
  perovskite system.  We systematically investigate the effects of
  oxygen octahedral rotations on the energies of RP phases in AFD
  perovskites (A$_{n+1}$B$_n$O$_{3n+1}$) for $n = 1\ldots30$,
  providing asymptotic results for $n \rightarrow \infty$ that give
  both the form of the interaction between stacking faults and the
  behavior of such stacking faults in isolation.  We find an
  inverse\hyp{}distance interaction between faults with a strength
  which varies by as much as a factor of two depending on the
  configuration of the octahedra.  We find that the strength of this
  effect can be sufficient to (a) stabilize or destabilize the RP
  phase with respect to dissociation into the bulk perovskite and the
  bulk A\hyp{}oxide and (b) affect the energy scales of the RP phase
  sufficiently to constrain the rotational states of the octahedra
  neighboring the stacking faults, even at temperatures where the
  octahedra in the bulk regions librate freely.  Finally, we present
  evidence that the importance of the octahedral rotations can be
  understood in terms of changes in the distances between oxygen ions
  on opposing sides of the RP stacking faults.
\end{abstract}




\maketitle

\section{Introduction}
\label{sec:intro}

Perovskites possess a vast range of scientifically interesting and
technologically important properties.  These materials are highly
valued for their di\-e\-lec\-tric\-i\-ty\cite{barrett1952dci,
hegenbarth1964dfd, samara1990ltd, shevlin2005aid},
ferro\-e\-lec\-tric\-i\-ty\cite{bednorz1984sax, cohen1992ofp,
haertling1999fch}, sem\-i\-con\-duc\-tiv\-i\-ty\cite{chan1981ns,
eror1982htd, frederikse1964eti, nakamura1971pst},
su\-per\-con\-duc\-tiv\-i\-ty\cite{schooley1964sss, schooley1965dot,
koonce1967stt}, cat\-a\-lyt\-ic ac\-tiv\-i\-ty\cite{arai1986ccm,
teraoka1990cap, pena2001csp}, and colossal
mag\-ne\-to\-re\-sist\-ance\cite{ramirez1997cm}.  Such physical
properties allow for the use of perovskites in diverse technological
applications, including tunneling semiconductor valves and magnetic
tunnel junctions in spintronics\cite{obata1999tma, deteresa1999rmo,
ziese2002emp}, dielectric insulators in dynamic random access
memory\cite{joshi1993sae, hwang1995dae}, thin films in graded
ferroelectric devices (GFDs)\cite{mantese1995ftf, vankeuls1997ysl},
and alternative gates in metal\hyp{}oxide\hyp{}semiconductor field
transistors (MOSFETs)\cite{mckee1998coo, edge2004mbo}.

\begin{figure*}
  \begin{centering}
    \subfloat[][{A$_2$BO$_4$}]{\includegraphics[width=0.4\columnwidth]{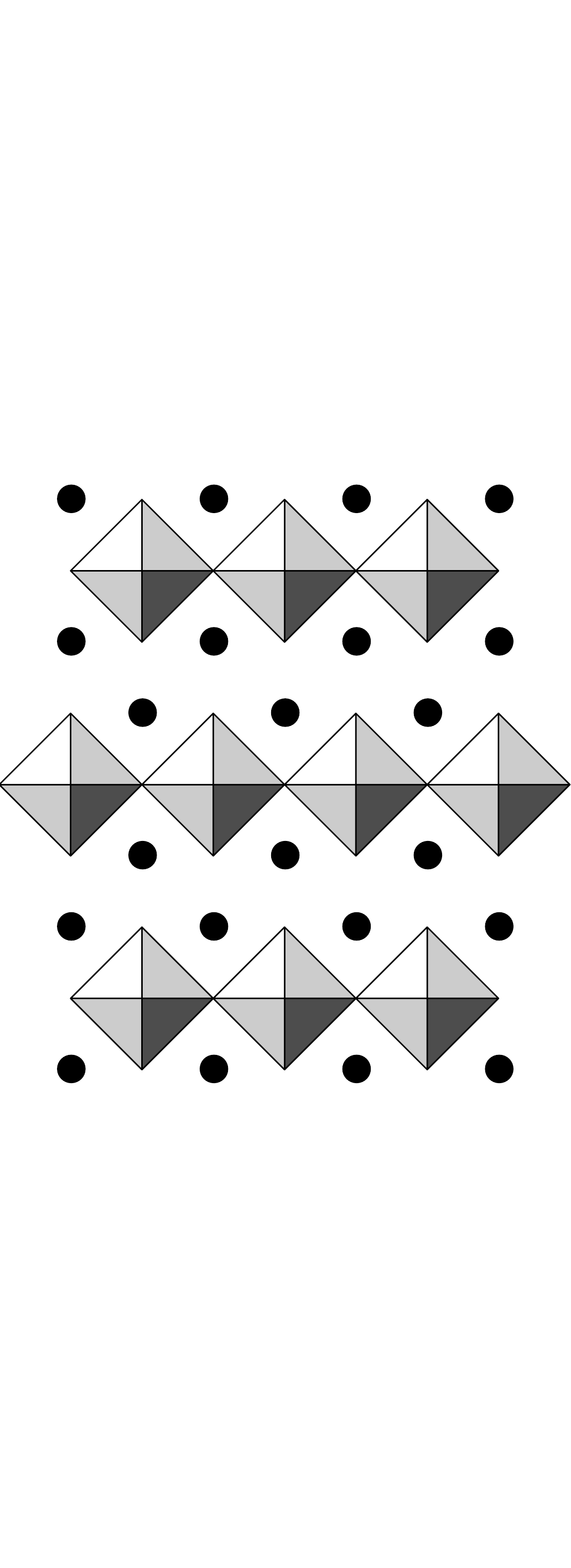}}
    \hspace{6ex}
    \subfloat[][{A$_3$B$_2$O$_7$}]{\includegraphics[width=0.4\columnwidth]{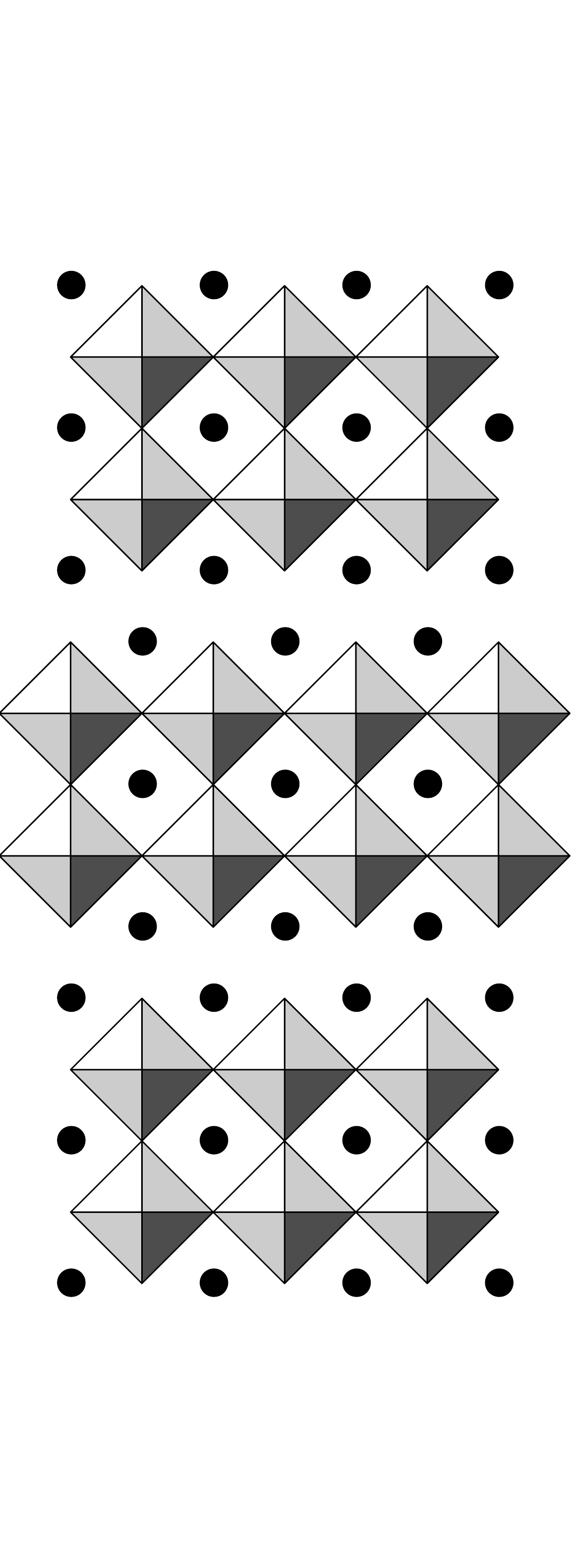}}
    \hspace{6ex}
    \subfloat[][{A$_4$B$_3$O$_{10}$}]{\includegraphics[width=0.4\columnwidth]{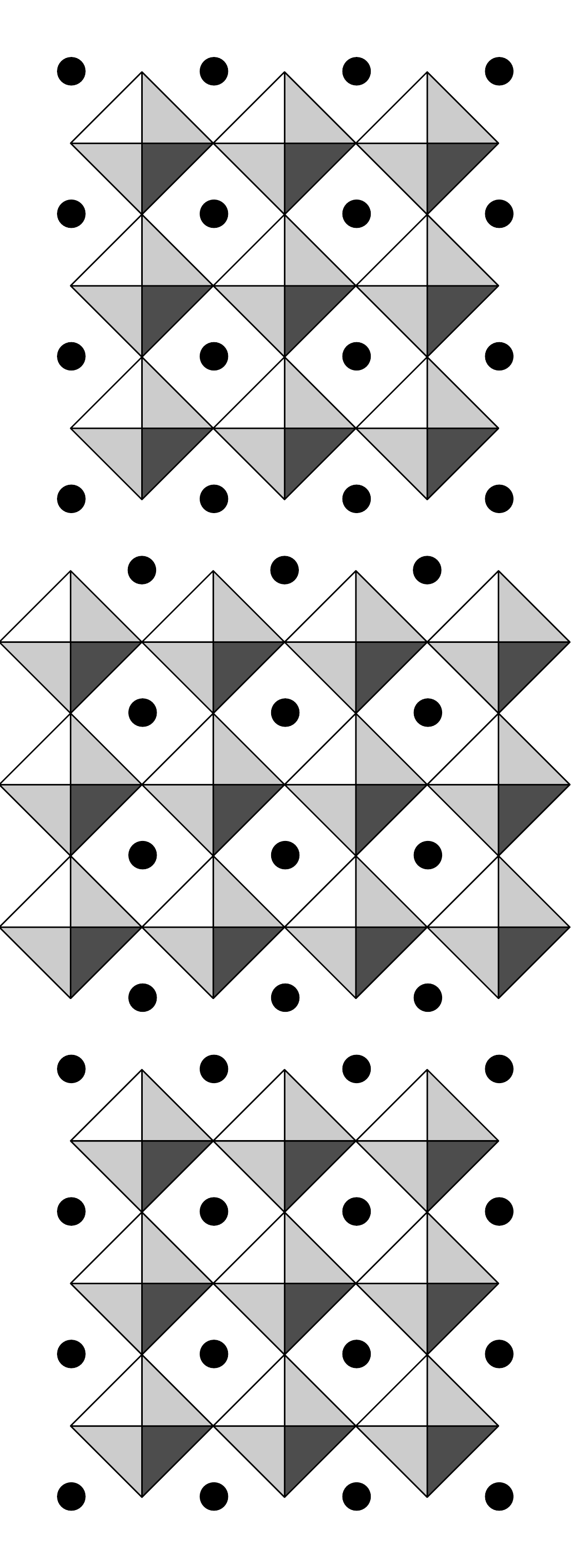}}
    \hspace{6ex}
    \subfloat[][{ABO$_3$}]{\includegraphics[width=0.4\columnwidth]{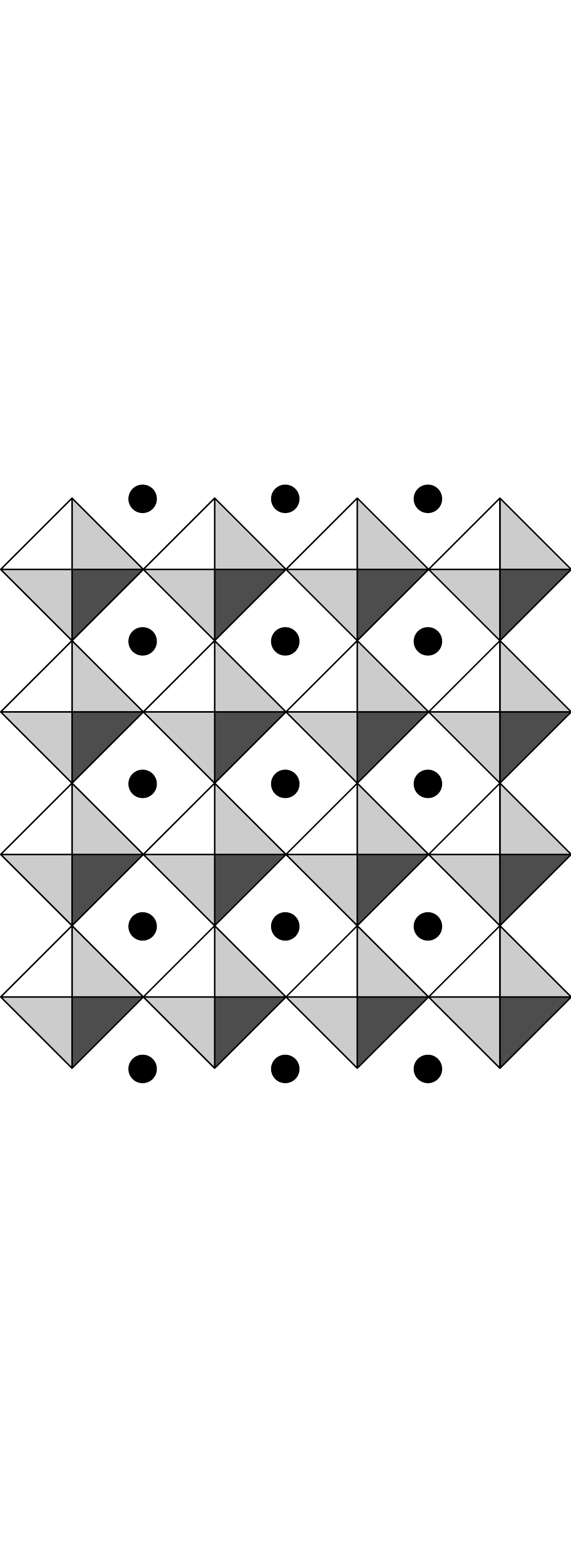}}
    \caption[Homologous compounds of the form
      A$_{n+1}$B$_n$O$_{3n+1}$]{Four members of series of homologous
      compounds of form A$_{n+1}$B$_n$O$_{3n+1}$, showing $n$ layers
      of bulk perovskite between stacking faults: (a) A$_2$BO$_4$ with
      $n = 1$, (b) A$_3$B$_2$O$_7$ with $n = 2$, (c)
      A$_4$B$_3$O$_{10}$ with $n = 3$, and (d) bulk ABO$_3$ with $n =
      \infty$.}
    \label{fig:seriesRP}
  \end{centering}
\end{figure*}

Cation stoichiometry has a large impact on the physics of these
materials, but is difficult\cite{yamamichi1994brd, taylor2003ios} to
control during film growth.  Cation non\hyp{}stoichiometries in the
perovskites can potentially take the form of point or planar defects,
with perovskites experimentally found to resist point defect formation
upon sufficient cation
non\hyp{}stoichiometry\cite{balachandran1982otd, smyth1985dao,
witek1984vot}.  In A$^{2+}$B$^{4+}$O$^{2-}_3$ materials (such as
SrTiO$_3$) with an excess of species A (or, more precisely, additional
AO to maintain charge neutrality), the resulting planar defects
produce a series of homologous compounds of the form
A$_{n+1}$B$_n$O$_{3n+1}$.  (Note that bulk ABO$_3$ corresponds to one
end member of this series, $n = \infty$, and bulk AO corresponds to
the other end member, $n = 0$.)  The A$_{n+1}$B$_n$O$_{3n+1}$
compounds, known as the Ruddlesden--Popper (RP)
phases\cite{ruddlesden1957nco, ruddlesden1958tcs}, reflect the
modification of the cation stoichiometry by the addition of an extra
unit of AO per $n$ units of bulk ABO$_3$ material.  Structurally, they
take the form of bulk ABO$_3$, separated every $n$ layers by the
insertion of excess $(001)$ planes of AO to create stacking faults of
the form $\ldots$/AO/BO$_2$/AO/AO/BO$_2$/AO/$\ldots\;$ in the normal
$\ldots$/AO/BO$_2$/AO/BO$_2$/AO$/\!\ldots\;$ stacking sequence of the
material.  Across each such stacking fault, the bulk perovskite slabs
are alternately displaced by in\hyp{}plane vectors of the form
$\frac{a_0}{2}[\pm1,\pm1,0\,]$.  Figure~\ref{fig:seriesRP} depicts
four members of the homologous series A$_{n+1}$B$_n$O$_{3n+1}$, with
$n = 1$, $2$, $3$, and $\infty$.

RP phases, specifically, manifest additional properties beyond those
of bulk perovskites; these include high
di\-e\-lec\-tric\-i\-ty\cite{kwestroo1959tsb, haeni2001egf},
ther\-mo\-e\-lec\-tric\-i\-ty\cite{lee2007tpr},
pho\-to\-ca\-tal\-y\-sis\cite{liang2008scs}, unconventional
su\-per\-con\-duc\-tiv\-i\-ty\cite{maeno1994slp}, quantum
crit\-i\-cal\-i\-ty\cite{grigera2001mft},
met\-a\-mag\-net\-ism\cite{perry2001mcf},
fer\-ro\-mag\-net\-ism\cite{cao1997fsr}, colossal
mag\-ne\-to\-re\-sist\-ance\cite{moritomo1996gmo}, and
x\nobreakdash-ray optic a\-dapt\-a\-bil\-i\-ty\cite{meyer2006rts}.
Despite the interest in such compounds, experimentalists have faced
difficulties growing RP phases for ``intermediate'' values of $n$ ($n
\gtrsim 4$)\cite{schlom1990mbe, klausmeier-brown1992eou, pena1994sam,
bader1998rip}.  For example, strontium titanate (SrTiO$_3$) compounds
with such stoichiometries tend not to form growths of a single uniform
RP phase, but rather materials with multiple regions, each with RP
phases of different values of $n$, so\hyp{}called
``intergrowths.''\cite{tilley1977aem, ramesh1990sds, schilling1993sak}
Indeed, conventional ceramic powder sintering has only produced
single\hyp{}phase RP samples for $n \le 3$: Sr$_2$TiO$_4$,
Sr$_3$Ti$_2$O$_7$, and Sr$_4$Ti$_3$O$_{10}$\cite{tilley1977aem,
mccarthy1969pei}.  More recently, successful creation of intermediate
members of the Sr$_{n+1}$Ti$_n$O$_{3n+1}$ series has been accomplished
with advanced techniques for epitaxial growth of thin\hyp{}films under
thermodynamical non\hyp{}equilibrium conditions, such as
sputtering\cite{taylor2003ios}, pulsed laser deposition
(PLD)\cite{iwazaki1999ass, iwazaki2000tso, shibuya2008slp}, and
molecular beam epitaxy (MBE)\cite{haeni2001egf, tanaka2000aco,
tian2001tem, fisher2007sli}, with characterization by
high\hyp{}resolution transmission electron microscopy
(HRTEM)\cite{tian2001tem} and x\nobreakdash-ray diffraction
(XRD)\cite{suzuki2000dsh}.  Such recent observations renew the urgency
for a careful theoretical study of RP phases with intermediate and
large values of $n$.

Previous theoretical work on RP phases of perovskites includes a
limited number of both \emph{ab initio} electronic structure
calculations and empirical potential studies.  A pioneering
empirical\hyp{}potential study\cite{udayakumar1988sao} in the 1980's
considered all values of $n \le 12$ for strontium titanate but did not
report sufficient precision to resolve differences among any of the
phases above $n = 1$.  A decade later, a subsequent
empirical\hyp{}potential study\cite{mccoy1997psa} was able to resolve
differences among phases but only considered phases for $n \le 4$.
Total energy electronic structure studies have also been carried out,
primarily for strontium titanate\cite{fennie2003sad, suzuki2001fps,
lebacq2006fps, noguera2000tir, weng2006eso, reshak2008ebs,
music2008eps} ($n \le 3$) but also for three additional perovskite
transition\hyp{}metal oxides\cite{weng2006eso} ($n = 2$ \emph{only}).
None of these studies have explored the significant interactions which
we find to exist between the rotational states of the oxygen octahedra
on opposite sides of the AO/AO stacking faults.

Many perovskites that form RP phases exhibit antiferrodistortive (AFD)
behavior in which the oxygen octahedra tend to rotate in an
alternating spatial pattern from their ideal orientations with a
relatively low energy scale, so that numerous structural phases exist
for the corresponding bulk materials.  Unless a full quantum
statistical treatment of the RP phases is considered in the
theoretical calculations, the bulk regions in these phases will tend
to relax to the zero\hyp{}temperature ground state within the model
used to describe the material.  One must therefore carefully and
explicitly consider two interrelated effects: (a) the interaction
between different possible antiferrodistortive reconstructions on
either side of the stacking faults and (b) the possibility that these
reconstructions will assume different orientations and thus prefer
lattice structures in potential conflict across the stacking fault.
If the different possible reconstructions of octahedral rotations are
not considered, the true ground state of the RP phases may be missed.
If the lattice vectors are not allowed to relax fully, the extraction
of the formation energy of the RP phases will suffer errors which
scale with the size of the bulk regions (i.e., linearly with $n$), as
we demonstrate below in Section~\ref{subsec:resultsFormation}.  The
aforementioned theoretical works thus have limited applicability, even
for the small values of $n$ which they consider.

Indeed, experiments reveal that these effects are known to be present
in RP phases of perovskites; for example, both
Sr$_{n+1}$Ti$_n$O$_{3n+1}$ and Ca$_{n+1}$Ti$_n$O$_{3n+1}$ exhibit
rotations of their oxygen octahedra, for $n = 2$ and $n = 3$ RP
phases\cite{elcombe1991sdc}.  Previous work has considered general
group\hyp{}theoretical studies of the space groups of possible
octahedral rotation patterns in RP
phases\cite{aleksandrov1994otp,hatch1989ptp}, but with restrictions
either to the $n = 1$ RP phase or to a limited number of possible
relative rotational orientations on opposite sides of the stacking
faults.

In this work, we provide a careful, comprehensive atomistic study of
the Ruddlesden--Popper phases of a physical model of an
antiferrodistortive perovskite, considering a wide range of $n$
(specifically, $n = 1\ldots30$), properly accounting for the full
lattice relaxation of the RP phases, and exploring all combinatorial
possibilities of different orientations of oxygen octahedra on
opposite sides of the AO/AO stacking faults.  In
Section~\ref{sec:background}, we introduce the class of materials
which we consider, those antiferrodistortive perovskites which form
intrinsic RP phases and belong to the Glazer system of the greatest
symmetry commonly found in nature: $a^-a^-a^-$\cite{glazer1972tco}.
Section~\ref{sec:classification} then presents an exhaustive catalogue
of the different possible structures for RP phases in perovskites
within this system.  In generating this catalogue, we introduce a
convenient symmetry algebra which allows one to quickly enumerate the
structures, and we use it to find a total of five
symmetry\hyp{}distinct possibilities (consistent with distorted phases
Nos.~$18$--$22$ which Hatch et al.\cite{hatch1989ptp}\ found for the
$n = 1$ RP phase after extensive searching with a computer program).
Section~\ref{sec:methods} then introduces the shell\hyp{}potential
model and the numerical techniques which we use.

Section~\ref{subsec:resultsFormation} explores the aforementioned five
distinct configurations for all RP phases for $n = 1\ldots30$.  We
find that the energies of RP phases are indeed quite sensitive to
octahedral rotations, sufficiently sensitive that unfavorable
configurations become unstable relative to phase separation into bulk
perovskite and bulk A\hyp{}oxide.  In fact, the effect is sufficiently
strong to suggest some intriguing possibilities.  For low densities of
stacking faults (high $n$), the rotational states of the octahedra
neighboring the faults might be constrained, even at temperatures
where the octahedra in the bulk librate freely, so that different
degrees of order are observed in the bulk and at the interfaces.  For
high densities of stacking faults (low $n$), this effect may increase
the transition temperatures associated with the octahedral rotations.

Section~\ref{subsec:resultsInteraction} considers interactions between
the stacking faults present in the RP phases.  For each configuration,
we examine the energy of the A$_{n+1}$B$_n$O$_{3n+1}$ RP phases as a
function of $n$, which directly measures the separation between
stacking faults.  We demonstrate that the interaction is quite
sensitive to the octahedral rotations, differing in strength by as
much as a factor of two depending on the rotations.  Next,
Section~\ref{subsec:resultsIsolated} examines the issue of the
asymptotic form of this interaction for $n \rightarrow \infty$.  We
find that the interaction between stacking faults varies as the
inverse of the distance between them, and we extract both the binding
energies of stacking faults and the formation energies of isolated
faults.  Again, we find that the interaction energy between faults is
highly sensitive to the different possible rotational states of the
oxygen octahedra and may even lead to ordering at the stacking faults
at temperatures where the bulk regions have lost their orientational
order.  This section then concludes with a proposal for a simple
physical mechanism to explain the strong dependence of the interfacial
energy on the rotational state of the octahedra: some configurations
result in movement of like\hyp{}charged neighboring oxygen ions
directly toward each other and thus are high in energy, whereas others
result in movement of oxygen ions past each other and thus are low in
energy.

Finally, in Section~\ref{subsec:resultsPoint}, exerting care in
tracking the chemical potentials of the various reference systems, we
verify that the formation of isolated planar stacking faults is
favored over the formation of point defects in our model material,
regardless of the rotational state of the oxygen octahedra.

\section{Background}
\label{sec:background}

In this work, we embark on a study of the generic effects of
octahedral rotations on RP phases.  For this first such study, we
shall focus on the most highly symmetric of the twenty\hyp{}three
possible Glazer systems\cite{glazer1972tco}.  There are two such
systems, denoted in Glazer notation as $a^+a^+a^+$ and $a^-a^-a^-$,
only the latter of which is commonly found in nature.  Examples of
$a^-a^-a^-$ perovskites include LaAlO$_3$\cite{glazer1972tco},
NdAlO$_3$\cite{geller1956csp, harley1973ptp},
CeAlO$_3$\cite{harley1973ptp}, BiFeO$_3$\cite{moreau1970adr},
LiNbO$_3$\cite{moreau1970adr}, LiTaO$_3$\cite{moreau1970adr},
PbZr$_{0.9}$Ti$_{0.1}$O$_3$ (PZT)\cite{glazer1975swo}, and many others
catalogued by Glazer\cite{glazer1972tco} and Megaw and
Darlington\cite{megaw1975gsr}.

The Glazer notation refers to the relative state of rotation of
neighboring oxygen octahedra in antiferrodistortive reconstructions of
the perovskite structure.  In actual fact, the motion of the octahedra
within such reconstructions is only approximately a rotation as the
oxygen atoms are confined to the faces of each cube; regardless, we
shall refer to their motion as rotational throughout.  Glazer
exhaustively catalogued all twenty\hyp{}three possible patterns of
these octahedral rotations (``tilts'' in his
terminology)\cite{glazer1972tco, glazer1975swo}, with each category
assigned an appropriate nomenclature to denote the axis of rotation
and the relative sign of successive rotations along that axis.  In
brief, in the $a^-a^-a^-$ system, all oxygen octahedra rotate either
clockwise or counterclockwise about a fixed trigonal axis in an
alternating checkerboard pattern in all three dimensions, resulting in
a cell\hyp{}doubling reconstruction with a $2 \times 2 \times 2$
supercell.  (Greater details appear in
Section~\ref{sec:classification}.)

The antiferrodistortive phase transition, associated with these
rotations, occurs due to the softening of the $\Gamma_{25}$ optical
phonons at the $R$ corner ($[111]$ zone boundary) of the Brillouin
zone, as famously studied experimentally\cite{unoki1967esr,
fleury1968spm, shirane1969lds} and theoretically\cite{pytte1969tsp,
feder1970toa} and reviewed extensively\cite{scott1974sms}.  The
transition is therefore also sometimes known as a Zone Boundary
Transition (ZBT) and was recognized even earlier as a means by which
crystals could double the size of their primitive
cells\cite{cochran1961cst}.  The rotation of the oxygen octahedra
accompany the softening of these phonons, with their rotation angles
serving as the order parameters\cite{muller1968csp}.  The degeneracy
of this phonon mode in the higher\hyp{}temperature cubic phase enables
the rotation of octahedra about different axes --- $\langle 100
\rangle$, $\langle 110 \rangle$, and $\langle 111 \rangle$ --- to
break the crystal's symmetry below its critical temperature.  While
the $a^-a^-a^-$ system results from rotations about the $\langle 111
\rangle$ axis, all three systems emerge from the same underlying
instability\cite{scott1974sms, axe1969zbp}, with anharmonic
interactions determining the resulting low\hyp{}temperature lattice
symmetry\cite{thomas1968spt}.  As the underlying physics is so
similar, we thus expect that perovskites in Glazer systems other than
$a^-a^-a^-$ will exhibit similar generic behaviors to the results
expounded below.

In seeking a model potential for our study, we searched the literature
for shell\hyp{}potential parameters among the $a^-a^-a^-$ perovskites.
Unfortunately, we discovered that the ground states of the available
models generally do not correctly match their corresponding Glazer
systems.\cite{lee2008pc} We were only able to identify a single
material, lanthanum aluminate (La$^{3+}$Al$^{3+}$O$^{2-}_3$), with a
ground state in the correct Glazer system.\cite{lee2008pc}
Unfortunately, lanthanum aluminate does not form \emph{intrinsic} RP
phases, since its composition as A$_{n+1}$B$_n$O$_{3n+1}$ would
violate basic charge balance; intrinsic RP formation requires
perovskites with an A$^{2+}$B$^{4+}$O$^{2-}_3$ chemical formula for
the additional A$^{2+}$O$^{2-}$ layer to be neutral.  Lanthanum
aluminate, however, can form \emph{extrinsic} RP phases by
incorporation of additional neutral layers of another perovskite,
strontium oxide: SrO$\cdot$La$_n$Al$_n$O$_{3n}$.

For simplicity of this initial theoretical study, we focus on
perovskites that form intrinsic RP phases, examples of which from the
$a^-a^-a^-$ system do indeed exist in nature (e.g.,
BaTbO$_3$\cite{glazer1972tco}).  However, we are not aware of shell
potentials for any of these materials.  On the other hand, we
discovered that the shell\hyp{}potential model commonly used for
strontium titanate (Sr$^{2+}$Ti$^{4+}$O$^{2-}_3$), which forms
intrinsic RP phases, does possess a ground state of the $a^-a^-a^-$
type.

Indeed, on a microscopic level, strontium titanate is very similar to
lanthanum aluminate.  In fact, early x\nobreakdash-ray
experiments\cite{lytle1964c, lytle1964xrd} erroneously predicted a
low\hyp{}temperature ($T \lesssim 35$~K) phase transition in strontium
titanate to a rhombohedral lattice (generally associated with
$a^-a^-a^-$ microscopic ordering), only to be corrected by subsequent
spectroscopic studies\cite{alefeld1969dmg, unoki1967esr}.  Strontium
titanate assumes its actual tetragonal ground\hyp{}state structure
through a famous cell\hyp{}doubling antiferrodistortive phase
transition near $105$~K\cite{vonwaldkirch1973fsn}, whose ``physical
origin is the same''\cite{muller1968csp} as that of lanthanum
aluminate.  Indeed, this transition is ``strikingly analogous in all
respects''\cite{scott1969rst, scott1970ers} to that of lanthanum
aluminate, although the energy scales are quite different.  (The
structural phase transition temperature for lanthanum aluminate is
$800$~K\cite{muller1968csp, howard2000npd}.)  Further connections
between these two perovskites are enabled by the classical concept of
the Goldschmidt tolerance factor\cite{goldschmidt1926uub},
$t=\frac{1}{\sqrt{2}}\frac{r_A+r_O}{r_B+r_O}$, a normalized ratio of
the radii of ABO$_3$ ions historically used to categorize perovskites
and predict their ground states\cite{galasso1969spp}.  In fact,
lanthanum aluminate has a tolerance factor within $1$\% of that of
strontium titanate (data compiled by Shannon\cite{shannon1976rei}).

In sum, the parameters for the shell potential commonly used for
strontium titanate describe a perovskite which intrinsically forms RP
phases within the desired $a^-a^-a^-$ Glazer system.  It is not
surprising, then, to discover empirical evidence for perovskites, such
as lanthanum aluminate, which closely resemble strontium titanate but
with $a^-a^-a^-$ octahedral rotations.  The standard shell potential
for strontium titanate can thus reasonably be utilized as a
parameterization for generic $a^-a^-a^-$ perovskites which form RP
phases.  Therefore, in interpreting our results, one should be mindful
that details, such as the values which we find for the various
quantities, may not apply to any specific perovskite.  However, the
general phenomena that we uncover, such as the enumeration of possible
octahedral configurations and the general form and scale of the
various interactions, may be taken as representative of the class of
$a^-a^-a^-$ perovskites in RP phases.  Moreover, since the underlying
physical mechanism is the same for other Glazer systems, many of the
phenomena which we uncover should be considered for the RP phases of
all AFD perovskites.

\section{Classification of octahedral rotations in RP phases}
\label{sec:classification}

\begin{figure}
  \begin{centering}
    \includegraphics[width=0.8\columnwidth]{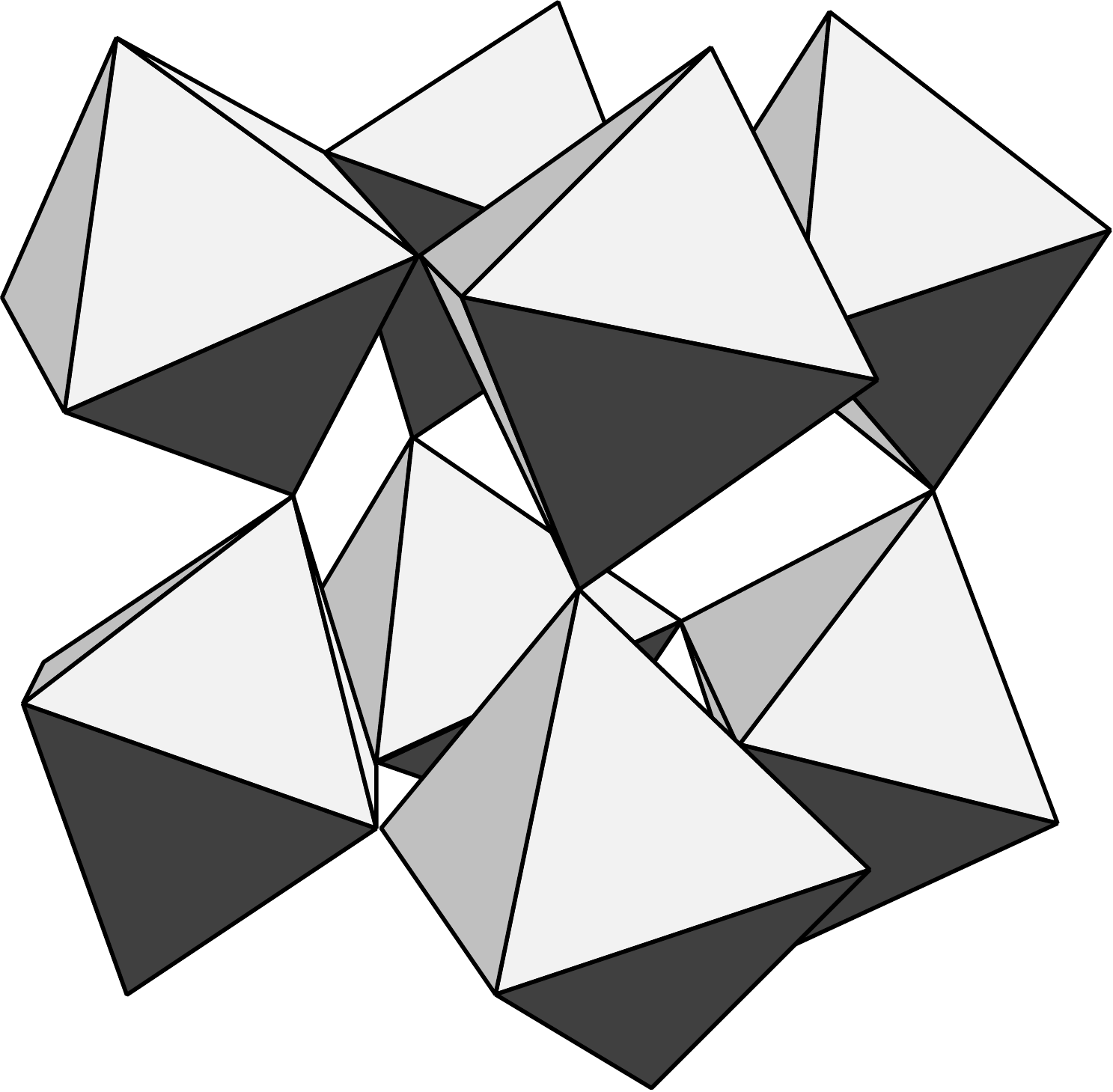}
    \caption[Perovskite reconstruction in $a^-a^-a^-$ Glazer
      notation]{Perovskite reconstruction in $a^-a^-a^-$ Glazer
      notation, showing rotations of oxygen octahedra in opposite
      directions in a cell\hyp{}doubling, alternating
      three\hyp{}dimensional $2 \times 2 \times 2$ checkerboard
      pattern.  (Degree of rotation exaggerated for illustrative
      purposes.)}
    \label{fig:reconstruction}
  \end{centering}
\end{figure}

We return now to a careful examination of the specifics of the
$a^-a^-a^-$ Glazer reconstruction, as visualized in
Figure~\ref{fig:reconstruction}.  The oxygen octahedra each rotate
around one of the eight possible $\langle 111 \rangle$ axes (expressed
in the coordinates of the closely related cubic structure), with
neighboring octahedra rotating in opposite directions in a
cell\hyp{}doubling, alternating three\hyp{}dimensional $2 \times 2
\times 2$ checkerboard pattern.  The crystal itself
responds to the presence of these trigonal rotations about a common
axis by stretching or compressing along that axis, forming a
rhombohedral Bravais lattice.  Selection of a specific oxygen
octahedron as reference then permits eight possible distinct bulk
configurations, each characterized and enumerated by the particular
choice of one of the eight possible trigonal rotation axes for that
specific octahedron.  These eight reconstructions can be characterized
either as $\pm[\pm1,\pm1,1\,]$, namely a selection of an overall $\pm$
sign and a choice of one of the four \emph{unsigned} $[\pm1,\pm1,1\,]$
rotation axes, or, alternatively, simply as one of eight \emph{signed}
$[\pm1,\pm1,\pm1]$ axes.

\begin{figure}
  \begin{centering}
    \includegraphics[width=0.6\columnwidth]{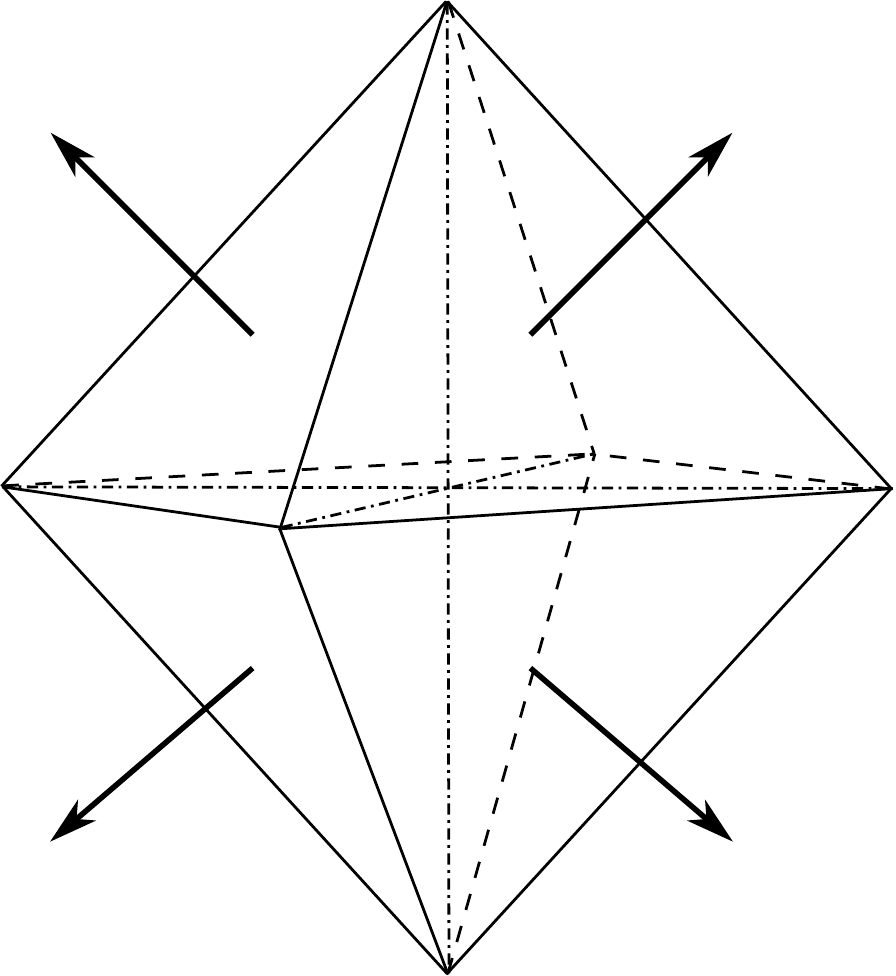}
    \caption[Trigonal rotation axes for oxygen octahedra]{``Positive''
      senses of the four (unsigned) $[\pm1,\pm1,1\,]$ rotation axes
      for oxygen octahedra ($\hat{e}_3$ axis out of the page).}
    \label{fig:octahedron}
  \end{centering}
\end{figure}

The former perspective is visualized in Figure~\ref{fig:octahedron}.
The four (unsigned) $[\pm1,\pm1,1\,]$ rotation axes designate the
reconstruction at an arbitrary origin of the crystal, where a final
choice of sign determines whether the rotation at the origin is either
clockwise (``positive'') or counterclockwise (``negative''), thereby
fully specifying the microscopic state of the crystal.  Note that a
change in this sign choice corresponds precisely to a rigid
translation of the crystal by a $[111]$ primitive translation vector
(to a position in the crystal with opposite sign of rotation, due to
the checkerboard pattern of the reconstruction).  Such a translation
would not be observable on a macroscopic level.

\subsection{Isolated stacking faults}
\label{subsec:classificationStackingFaults}

The RP phases observed in experiments\cite{tilley1977aem,
mccarthy1969pei, mccarthy1969tse} consist of superlattices of bulk
perovskite slabs separated by the insertion of excess $(001)$ AO
planes to create stacking faults of the form
$\ldots$/AO/BO$_2$/AO/AO/BO$_2$/AO/$\!\ldots$ in the normal
$\ldots$/AO/BO$_2$/AO/BO$_2$/AO/$\!\ldots\;$ stacking sequence.
Across each such stacking fault, the bulk perovskite slabs are
alternately displaced by in\hyp{}plane vectors of the form
$\frac{a_0}{2}[\pm1,\pm1,0\,]$.  To our knowledge, however, no one has
yet explored the effects of different combinations of possible
symmetry\hyp{}related bulk reconstructions on opposite sides of the
repeated AO planes.

To enumerate the distinct possible configurations for such stacking
faults, we restrict our consideration to cases where the material on
either side of the stacking fault possesses a specific bulk
reconstruction throughout.  We first focus on the reconstruction of
the bulk material on the side ``below'' (at lower values for
$\hat{e}_3$) the AO/AO stacking fault, which we shall denote as side
$\mathfrak{A}$.  As discussed above, the microscopic configuration of
bulk material on side $\mathfrak{A}$ can be fully specified by noting
the rotation axis of the octahedron at an arbitrary, but from then
onward fixed, origin $O_\mathfrak{A}$.  The rotation axis of this
particular reference oxygen octahedron can then assume any of eight
choices among the $\langle 111 \rangle$ axes, which we can regard as a
selection of an overall $\pm$ sign (``positive'' or ``negative'') and
a choice of one of the four (unsigned) $[\pm1,\pm1,1\,]$ rotation
axes.  This specification then determines the entire microscopic
structure of the bulk material in side $\mathfrak{A}$, according to
the $2 \times 2 \times 2$ three\hyp{}dimensional checkerboard pattern
of alternating signs for the rotation axes.  For the present purpose,
we choose always to select the reference octahedron $O_\mathfrak{A}$
from among those octahedra immediately neighboring the stacking fault
which have a positive rotation in the above sense of the choice of
overall $\pm$ sign.

\begin{figure}
  \begin{centering}
    \subfloat[][{$ + O_\mathfrak{B}$}]{\includegraphics[width=0.4\columnwidth]{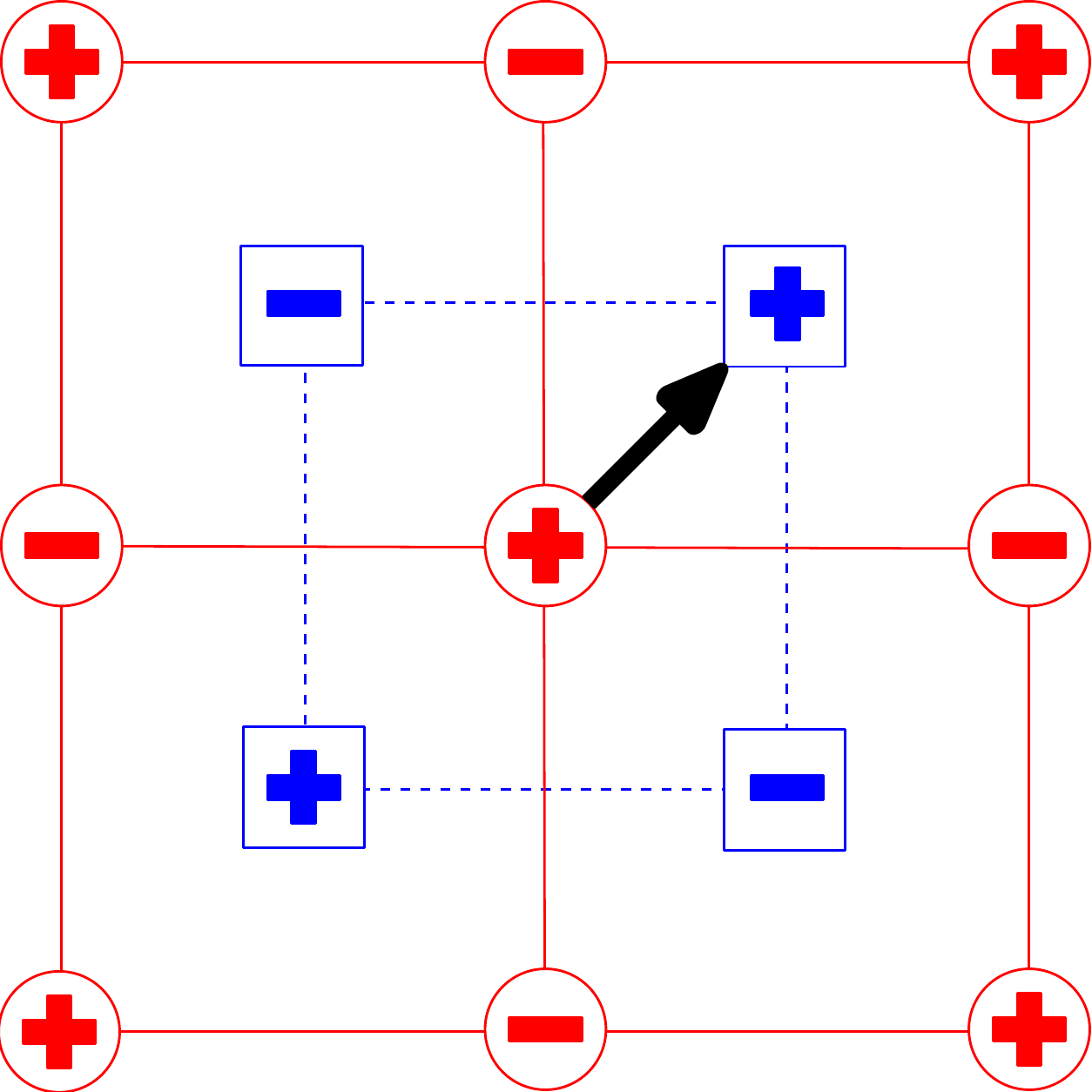}}
    \hspace{6ex}
    \subfloat[][{$ - O_\mathfrak{B}$}]{\includegraphics[width=0.4\columnwidth]{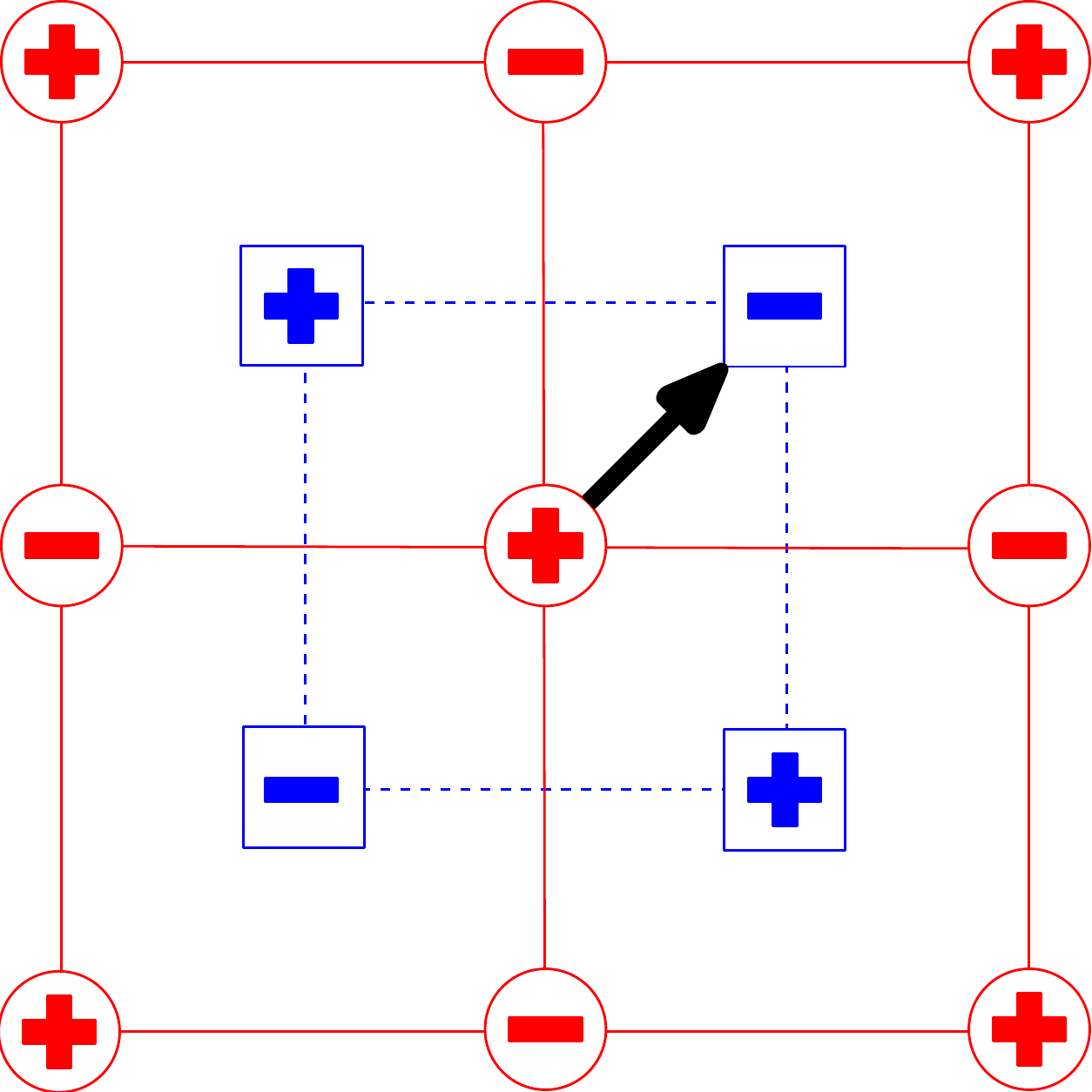}}\\
    \caption[Sign pattern of octahedral rotations neighboring stacking
      fault]{Sign pattern of octahedral rotations on opposite sides of
      stacking fault: octahedra in lower layer $\mathfrak{A}$
      (\textcolor{red}{$\mspace{0.2mu}\Circle\mspace{0.2mu}$}) and
      upper layer $\mathfrak{B}$ (\textcolor{blue}{$\Box$}) and
      standardized displacement $\mathbf{D}$
      ($\boldsymbol{\rightarrow}$).  Panels~(a) and~(b) depict the two
      distinct choices for sign of rotation ($+$ and $-$,
      respectively) of $O_\mathfrak{B}$.}
    \label{fig:stackedSigns}
  \end{centering}
\end{figure}

Next, as noted above, the bulk perovskite on the other side,
$\mathfrak{B}$, of the stacking fault is displaced, in general, by a
vector $\frac{a_0}{2}[\pm1,\pm1,0\,]$ relative to side $\mathfrak{A}$.
Temporarily disregarding the rotational state of all octahedra, we
note that all four of these displacements are equivalent, since the
octahedra on $\mathfrak{B}$ are positioned at the centers of squares
formed by the octahedra on $\mathfrak{A}$.  (See
Figure~\ref{fig:stackedSigns}.)  This equivalence (apart from
rotational states of the octahedra) allows us to choose a standardized
displacement $\mathbf{D} \equiv \frac{a_0}{2}[+1,+1,0\,]$ from
$O_\mathfrak{A}$ to select the origin $O_\mathfrak{B}$.  (More
specifically, $O_\mathfrak{B} = O_\mathfrak{A} + \mathbf{D} + \zeta
a_0 \hat{e}_3$, where $\zeta \approx 3/2$ represents the vertical
displacement between octahedra on opposite sides of the stacking
fault, and $\hat{e}_3$ is a unit vector in the $[001]$ direction.)

The rotation state of $O_\mathfrak{B}$ then completely specifies the
entire microscopic structure of the bulk material in side
$\mathfrak{B}$.  This rotation state can be specified as a selection
of an overall $\pm$ sign (``positive'' or ``negative'') and a choice
of one of the four (unsigned) $[\pm1,\pm1,1\,]$ rotation axes.
Figure~\ref{fig:stackedSigns} illustrates exactly these two possible
choices of sign, where the reference octahedron on side $\mathfrak{B}$
is either positive (Figure~\ref{fig:stackedSigns}(a)) or negative
(Figure~\ref{fig:stackedSigns}(b)).  Generically below, we shall refer
to these two possible sign patterns in the stacking\hyp{}fault
configuration as ``$+$'' and ``$-$'', respectively, reflecting the
overall sign of the rotation of $O_\mathfrak{B}$.  The combination of
these two sign patterns and the four possible unsigned rotation axes
(not indicated in Figure~\ref{fig:stackedSigns}) for each origin,
$O_\mathfrak{A}$ and $O_\mathfrak{B}$, leads to a total of $2 \times
(4 \times 4) = 32$ distinct possible configurations for this stacking
fault.

Next, we consider equivalence of stacking faults under application of
$C_{4z}$ symmetry.  As discussed above, the rotation axis of
$O_\mathfrak{A}$ can always be selected to be among the four
(unsigned) $[\pm1,\pm1,1\,]$ axes.  Because $C_{4z}$ symmetries
interconvert all four of these axes ($C_{4z} \circ [111] =
[\bar{1}11]$, $C_{4z}^2 \circ [111] = [\bar{1}\bar{1}1]$, $C_{4z}^3
\circ [111] = [1\bar{1}1]$), an overall rotation of the coordinate
system can be found to make the $O_\mathfrak{A}$ rotation about the
$[111]$ axis.  We may thus define the rotation axis of
$O_\mathfrak{A}$ to always be the (unsigned) $[+1,+1,1\,]$ axis.  This
then narrows the phase space of thirty\hyp{}two configurations listed
above to now only $2 \times (1 \times 4) = 8$ distinct configurations.

These eight distinct configurations can be enumerated using two
related nomenclatures (seen in Table~\ref{tab:enumeratedConfigs}), one
which is algebraically explicit and suitable for symmetry arguments
and the other which is more compact and convenient for communication.
In the former case of the \emph{algebraic specifier}, we enumerate
each configuration by specification of the rotation states of each
reference octahedron, $O_\mathfrak{A}$ and $O_\mathfrak{B}$,
expressing the rotation states in terms of (signed) $[\pm1,\pm1,\pm1]$
vectors.  For example, if $O_\mathfrak{A}$ is in rotation state
$[111]$ and $O_\mathfrak{B}$ is in rotation state $[11\bar{1}]$, we
write $(111 \Longrightarrow 11\bar{1})$.  For a more compact notation,
we can take, without loss of generality as shown above, the rotation
state of $O_\mathfrak{A}$ to always be $[111]$.  The unsigned rotation
axis of $O_\mathfrak{B}$ is then related to that of $O_\mathfrak{A}$
by one of the four $C_{4z}$ rotations of angles \{$0$,
$\frac{\pi}{2}$, $\pi$, $\frac{\smash{3}\pi}{2}$\}.  To denote the
configuration of the stacking fault, we then append the remaining
choice of $\pm$ sign for $O_\mathfrak{B}$ to its rotation angle to
produce a compound \emph{rotation\hyp{}sign specifier}.  For example,
the $(111 \Longrightarrow 11\bar{1})$ configuration from the preceding
example may be more compactly written as $[\pi^-]$ (since $-C_{4z}^2
\circ [111] = -[\bar{1}\bar{1}1] = [11\bar{1}]$).  Finally,
Table~\ref{tab:enumeratedConfigs} enumerates all eight configurations
according to both their rotation\hyp{}sign and algebraic specifiers.

\begin{table}
  \setlength{\doublerulesep}{0\doublerulesep}
  \setlength{\tabcolsep}{5\tabcolsep}
  \begin{tabular}{cc}
    \hline\hline\\[-1.5ex]
    Rotation-sign Specifier & Algebraic Specifier\\[0.5ex]
    \hline\\[-1.5ex]
    $[0^+]$                      & $(111 \Longrightarrow 111)$\\[0.3ex]
    $[0^-]$                      & $(111 \Longrightarrow \bar{1}\bar{1}\bar{1})$\\[0.3ex]
    $[\frac{\pi}{2}^+]$          & $(111 \Longrightarrow \bar{1}11)$\\[0.3ex]
    $[\frac{\pi}{2}^-]$          & $(111 \Longrightarrow 1\bar{1}\bar{1})$\\[0.3ex]
    $[\pi^+]$                    & $(111 \Longrightarrow \bar{1}\bar{1}1)$\\[0.3ex]
    $[\pi^-]$                    & $(111 \Longrightarrow 11\bar{1})$\\[0.3ex]
    $[\frac{\smash{3}\pi}{2}^+]$ & $(111 \Longrightarrow 1\bar{1}1)$\\[0.3ex]
    $[\frac{\smash{3}\pi}{2}^-]$ & $(111 \Longrightarrow \bar{1}1\bar{1})$\\[0.5ex]
    \hline\hline
  \end{tabular}
  \caption[Enumeration of interfacial stacking faults]{Enumeration of
    interfacial stacking faults remaining after application of
    $C_{4z}$ and translational symmetry, labeled by either relative
    rotation and sign of the bulk regions on opposite sides of the
    stacking fault (\emph{rotation\hyp{}sign specifier}) or rotation
    state of the two reference octahedra (\emph{algebraic
    specifier}).}
  \label{tab:enumeratedConfigs}
\end{table}

\begin{figure}
  \begin{centering}
    \subfloat[][{$[\frac{\pi}{2}^+]$}]{\includegraphics[width=0.4\columnwidth]{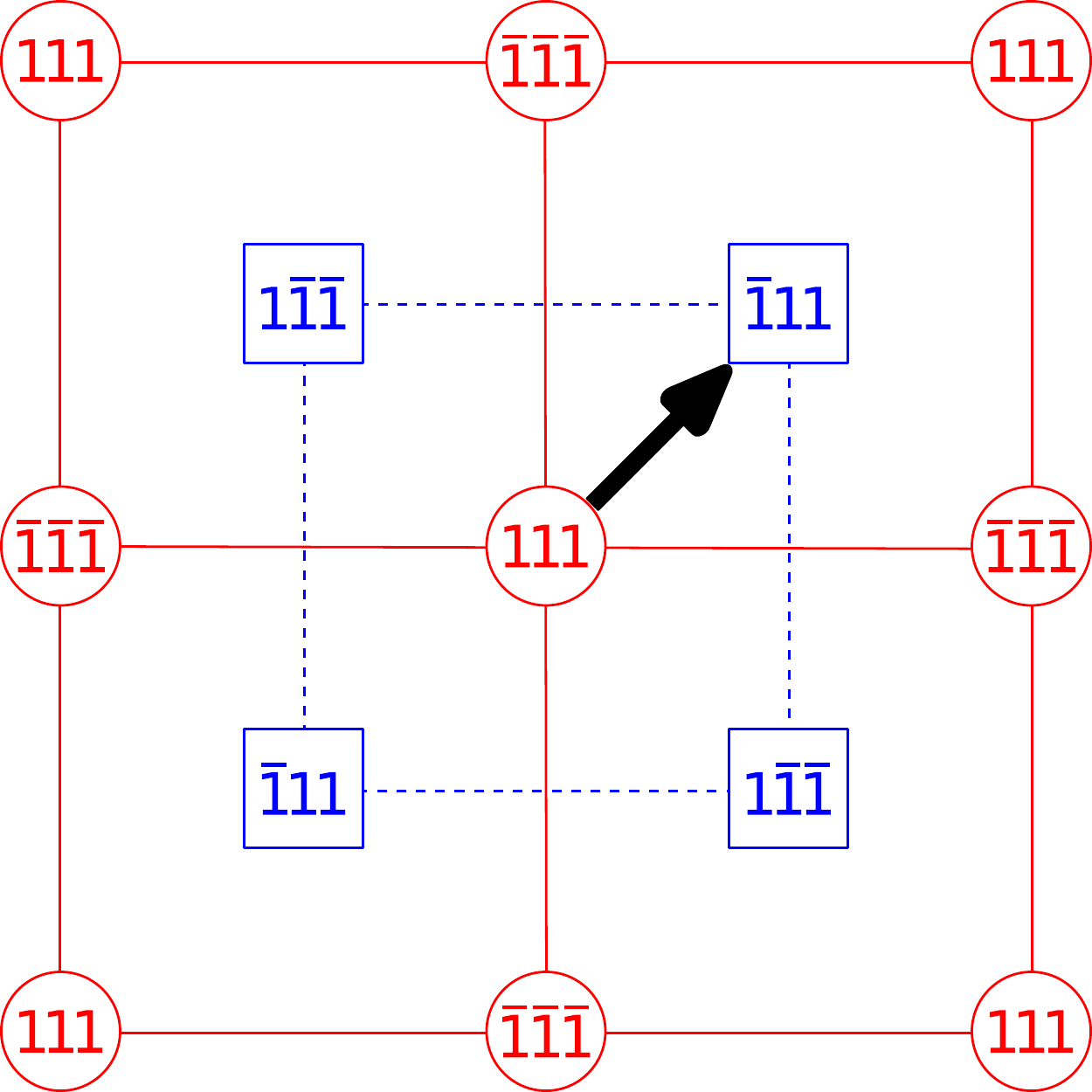}}
    \hspace{6ex}
    \subfloat[][{$[\frac{\pi}{2}^-]$}]{\includegraphics[width=0.4\columnwidth]{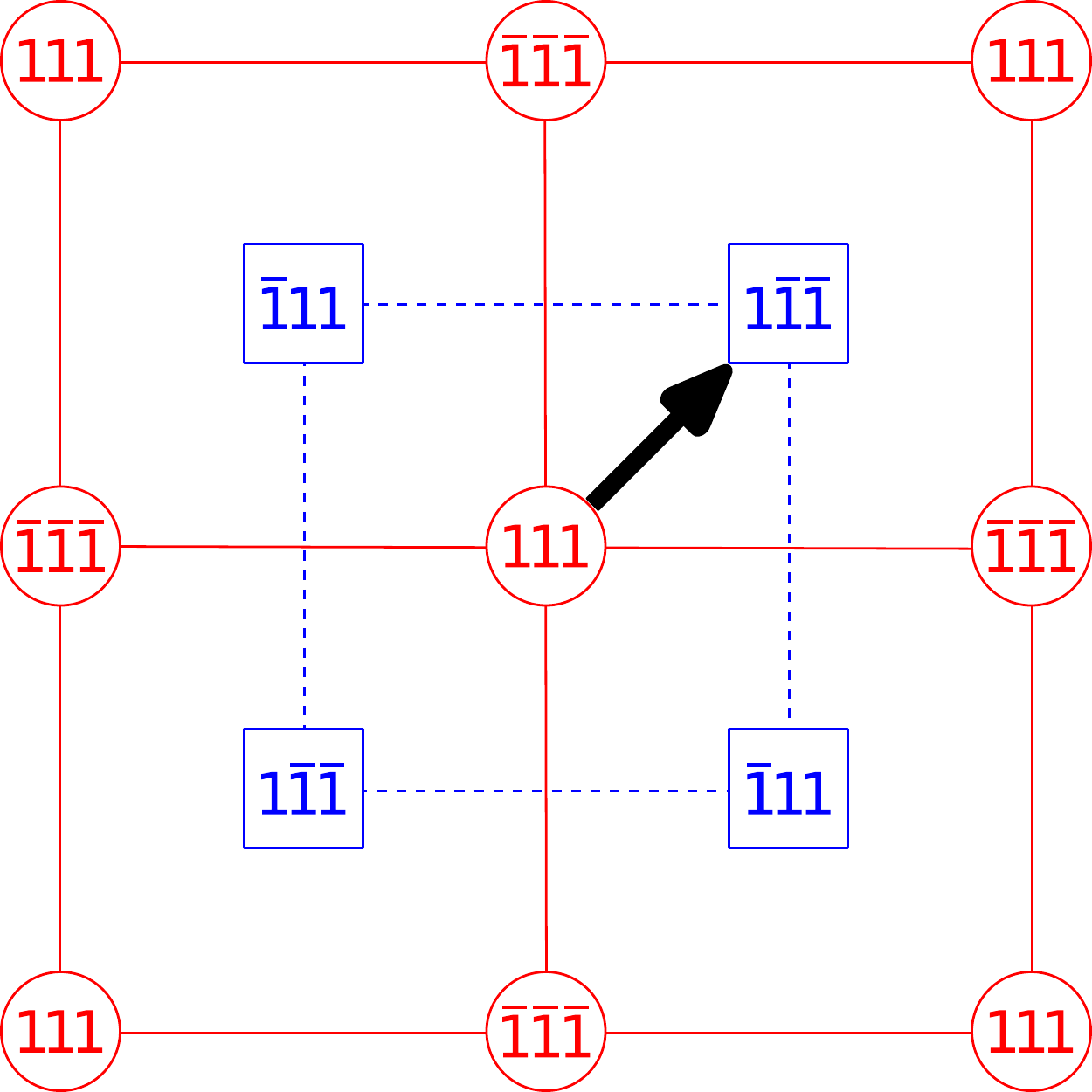}}\\
    \subfloat[][{$C_{2x} \circ [\frac{\pi}{2}^+]$}]{\includegraphics[width=0.4\columnwidth]{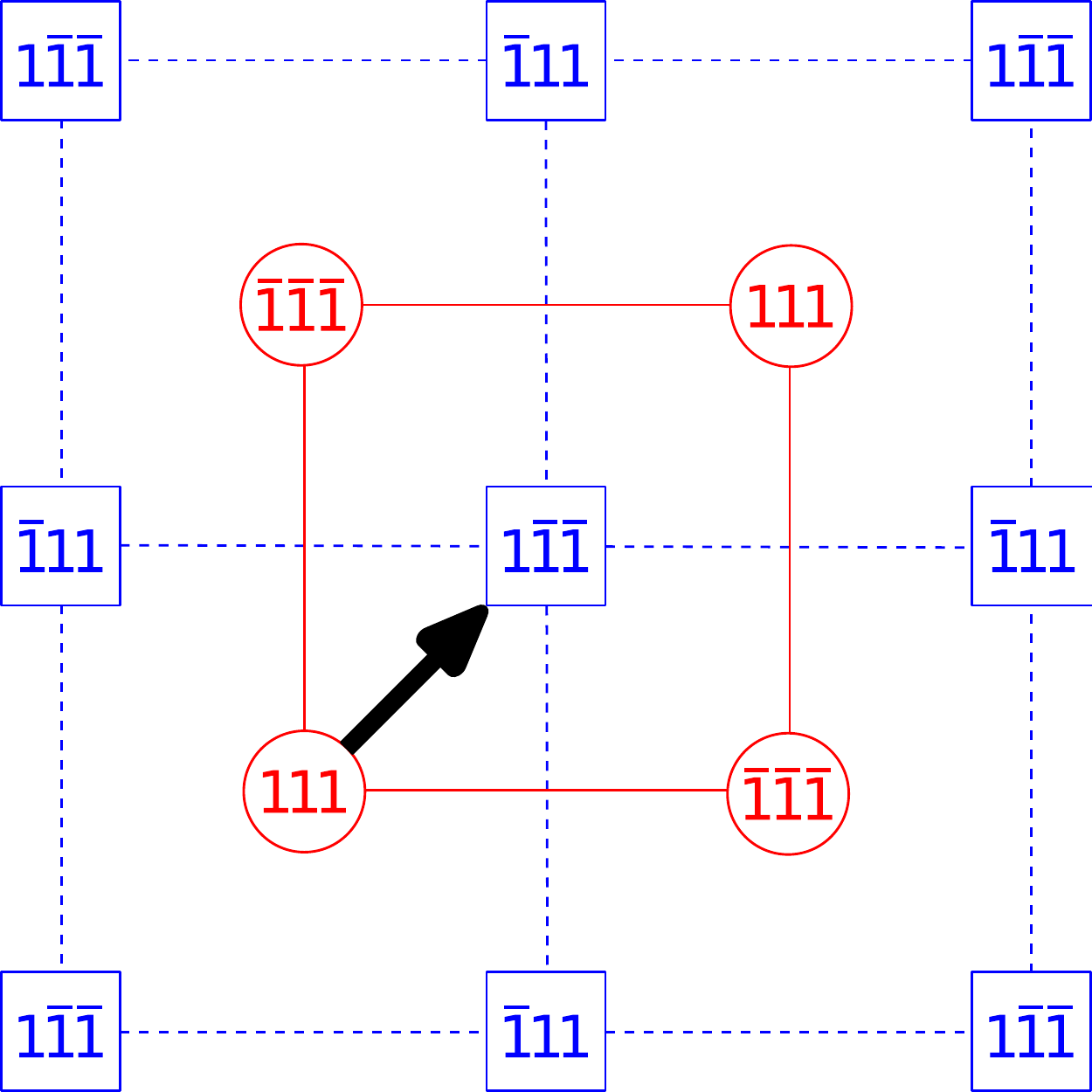}}
    \caption[Full specification of octahedral rotations neighboring
      stacking fault]{Full specification of octahedral rotations on
      opposite sides of stacking fault: octahedra in lower layer
      $\mathfrak{A}$
      (\textcolor{red}{$\mspace{0.2mu}\Circle\mspace{0.2mu}$}) and
      upper layer $\mathfrak{B}$ (\textcolor{blue}{$\Box$}) and
      standardized displacement $\mathbf{D}$
      ($\boldsymbol{\rightarrow}$), with explicit (signed)
      $[\pm1,\pm1,\pm1]$ rotation axes to specify full configuration.
      Panels~(a) and~(b) depict different configurations, with
      Panel~(c) showing the application of $C_{2x}$ rotation to (a) to
      produce a configuration equivalent to (b).}
    \label{fig:stackedVecs}
  \end{centering}
\end{figure}

The above enumerated eight stacking\hyp{}fault configurations can be
reduced yet further by considering the effects of additional, more
subtle, symmetry operations --- specifically, the $C_{2x}$, $C_{2y}$,
and $C_{4z}$ rotations and inversion $I$.  To demonstrate the effects
of such symmetry operations, we first consider two of the above eight
stacking\hyp{}fault configurations, $[\frac{\pi}{2}^+]$ in
Figure~\ref{fig:stackedVecs}(a) and $[\frac{\pi}{2}^-]$ in
Figure~\ref{fig:stackedVecs}(b).  From the figure, these two
configurations clearly differ only in the overall choice of sign
pattern for the upper layer $\mathfrak{B}$
(\textcolor{blue}{$\Box$}'s).  We now use symmetry arguments to
demonstrate that these two stacking\hyp{}fault configurations are
actually equivalent, related simply by a $C_{2x}$ rotation.

The general strategy which we shall employ is to exploit the fact that
the algebraic specifier not only determines the rotation state of all
of the octahedra in the crystal but also transforms in relatively
simple ways under the application of symmetry operations to the
overall crystal.  Figure~\ref{fig:stackedVecs}(c) illustrates the
state of the crystal in Figure~\ref{fig:stackedVecs}(a) after the
application of a $C_{2x}$ symmetry about a \emph{specific} axis.  To
ensure that the basic crystalline structure maps back onto itself,
this axis is chosen to pass at equal distances to the two neighboring
AO layers in the stacking fault and through the point immediately
above the $O_\mathfrak{A}$ reference octahedron (base of the
standardized displacement $\mathbf{D}$ in
Figure~\ref{fig:stackedVecs}(a)).  The application of $C_{2x}$ effects
the following changes to the configuration in
Figure~\ref{fig:stackedVecs}(a):
\begin{enumerate}
\item the position of $O_\mathfrak{A}$ maps from the bottom layer
  $\mathfrak{A}$ to the top layer $\mathfrak{B}$ (this changes the
  central \textcolor{red}{$\Circle$} in
  Figure~\ref{fig:stackedVecs}(a) into the central
  \textcolor{blue}{$\Box$} in Figure~\ref{fig:stackedVecs}(c));
\item the rotation state of $O_\mathfrak{A}$ maps from $[111]$ to
  $[1\bar{1}\bar{1}]$;
\item the position of the $\mathfrak{B}$\hyp{}side octahedron
  (\textcolor{blue}{$\Box$}) in the lower\hyp{}right quadrant of
  Figure~\ref{fig:stackedVecs}(a) maps to the $\mathfrak{A}$\hyp{}side
  octahedron (\textcolor{red}{$\Circle$}) in the upper\hyp{}right
  quadrant of Figure~\ref{fig:stackedVecs}(c);
\item the rotation state of this latter octahedron maps from
  $[1\bar{1}\bar{1}]$ to $[111]$.
\end{enumerate}
To determine the algebraic specifier for the final configuration in
Figure~\ref{fig:stackedVecs}(c), we now identify an octahedron on the
lower side of the stacking fault which has a positive rotation from
among the (unsigned) $[\pm1,\pm1,1\,]$ axes (base of the standardized
displacement $\mathbf{D}$ in Figure~\ref{fig:stackedVecs}(c)).  The
rotation state of this octahedron, combined with that of the
corresponding reference octahedron on the upper side, defines the
algebraic specifier for the resulting stacking fault.  For this
particular case, it is evident from the figure that $C_{2x} \circ
[\frac{\pi}{2}^+] = (111 \Longrightarrow 1\bar{1}\bar{1})$.  Reference
to Table~\ref{tab:enumeratedConfigs} then identifies $(111
\Longrightarrow 1\bar{1}\bar{1}) = [\frac{\pi}{2}^-]$, so that $C_{2x}
\circ [\frac{\pi}{2}^+] = [\frac{\pi}{2}^-]$, as direct comparison of
Figures~\ref{fig:stackedVecs}(b) and~(c) confirms.  Thus, we conclude
that $[\frac{\pi}{2}^+]$ and $[\frac{\pi}{2}^-]$ are identical and
related by the $C_{2x}$ rotation.

To develop an algebra for the action of symmetry operations on
arbitrary stacking\hyp{}fault configurations, we consider the effects
of such operations on general stacking\hyp{}fault configurations,
$(abc \Longrightarrow \alpha\beta\gamma)$.  For the case of $C_{2x}$,
we note in preparation that the action of $C_{2x}$ on any vector
$(xyz)$ is $C_{2x} \circ (xyz) = (x\bar{y}\bar{z})$.  Then, as above,
the operation of $C_{2x}$ transforms the generic algebraic specifier
$(abc \Longrightarrow \alpha\beta\gamma)$ as follows:
\begin{enumerate}
\item it swaps $abc$ and $\alpha\beta\gamma$;
\item it maps $abc$ to $a\bar{b}\bar{c\vphantom{b}}$;
\item it changes the sign of $\alpha\beta\gamma$;
\item it maps $\alpha\beta\gamma$ to
  $\alpha\bar{\beta}\bar{\gamma\vphantom{\beta}}$
\end{enumerate}
In short,
\begin{equation*} \label{eqn:sampleSym}
  C_{2x} \circ (abc \Longrightarrow \alpha\beta\gamma) = (\bar{\alpha\vphantom{\beta}}\beta\gamma \Longrightarrow a\bar{b\vphantom{\beta}}\bar{c\vphantom{\beta}}).
\end{equation*}
Table~\ref{tab:symOps} summarizes the effects, similarly determined,
on stacking\hyp{}fault configurations of a number of useful symmetry
operations: rotations $C_{2x}$, $C_{2y}$, $C_{2z}$, and $C_{4z}$ and
inversion $I$.  (Note that inversion, through the midpoint of the
vector connecting the two reference octahedra $O_\mathfrak{A}$ and
$O_\mathfrak{B}$, requires special care: it does not change the
direction of pseudo\hyp{}vectors such as the rotation axes of the
different octahedra but simply interchanges the two sides,
$\mathfrak{A}$ and $\mathfrak{B}$.)

\begin{table}
  \setlength{\doublerulesep}{0\doublerulesep}
  \setlength{\tabcolsep}{5\tabcolsep}
  \begin{tabular}{ccc}
    \hline\hline\\[-1.5ex]
    \multirow{2}{*}[-0.5ex]{Symmetry} & \multicolumn{2}{c}{Algebraic Specifier}\\[0.5ex]
    \cline{2-3}\\[-1.5ex]
    & Original & Final\\[0.5ex]
    \hline\\[-1.5ex]
    $C_{2x}$ & $(abc \Longrightarrow \alpha\beta\gamma)$ & $(\alpha\bar{\beta}\bar{\gamma\vphantom{\beta}} \Longrightarrow \mathrlap{\:\bar{a\vphantom{\beta}}bc} \hphantom{\alpha\beta\gamma})$\\[0.3ex]
    $C_{2y}$ & $(abc \Longrightarrow \alpha\beta\gamma)$ & $(\bar{\alpha\vphantom{\beta}}\beta\bar{\gamma\vphantom{\beta}} \Longrightarrow  \mathrlap{\:a\bar{b\vphantom{\beta}}c} \hphantom{\alpha\beta\gamma})$\\[0.3ex]
    $C_{2z}$ & $(abc \Longrightarrow \alpha\beta\gamma)$ & $(\mathrlap{\:\bar{a\vphantom{\beta}}\bar{b\vphantom{\beta}}c} \hphantom{\alpha\beta\gamma} \Longrightarrow \bar{\alpha\vphantom{\beta}}\bar{\beta\vphantom{\beta}}\gamma)$\\[0.3ex]
    $C_{4z}$ & $(abc \Longrightarrow \alpha\beta\gamma)$ & $(\mathrlap{\:\bar{b\vphantom{\beta}}ac} \hphantom{\alpha\beta\gamma} \Longrightarrow \beta\bar{\alpha\vphantom{\beta}}\bar{\gamma\vphantom{\beta}})$\\[0.3ex]
    $I$ & $(abc \Longrightarrow \alpha\beta\gamma)$ & $(\alpha\beta\gamma \Longrightarrow \mathrlap{\:abc} \hphantom{\alpha\beta\gamma})$\\[0.5ex]
    \hline\hline
  \end{tabular}
  \caption[Rotation and inversion symmetry operations]{Rotation ($C$)
    and inversion ($I$) symmetry operations applied to general
    stacking\hyp{}fault configuration.}
  \label{tab:symOps}
\end{table}

We are now able to demonstrate further symmetry reduction of the
stacking\hyp{}fault configurations.  Table~\ref{tab:enumeratedConfigs}
lists all eight possible stacking fault configurations, providing both
rotation\hyp{}sign and algebraic specifiers for each.  Application of
the symmetry operations from Table~\ref{tab:symOps} to these
configurations, and subsequent translation $T$ by $a_0 [100]$ if
necessary to make $O_\mathfrak{A}$ of rotation state $[111]$,
generates the following relationships,
\begin{subequations} \label{eqn:allSyms}
  \begin{align}
    T \circ C_{2y} \circ [\tfrac{\smash{3}\pi}{2}^+] &= [\tfrac{\smash{3}\pi}{2}^-]\\
    T \circ C_{2x} \circ [\tfrac{\pi}{2}^+] &= [\tfrac{\pi}{2}^-]\\
    C_{4z} \circ I \circ [\tfrac{\smash{3}\pi}{2}^+] &= [\tfrac{\pi}{2}^-].
  \end{align}
\end{subequations}
The above symmetry operations, which each involve interchange of the
$\mathfrak{A}$ and $\mathfrak{B}$ sides of the interface, demonstrate
equivalence among the set of four stacking faults,
\{$[\frac{\pi}{2}^+]$, $[\frac{\pi}{2}^-]$,
$[\frac{\smash{3}\pi}{2}^+]$, $[\frac{\smash{3}\pi}{2}^-]$\}.  Thus,
only five unique stacking\hyp{}fault configurations remain, $[0^+]$,
$[0^-]$, $[\frac{\pi}{2}^+]$, $[\pi^+]$, and $[\pi^-]$.  Comparison of
the algebraic specifiers of these five configurations, from
Table~\ref{tab:enumeratedConfigs} above, with the
computer\hyp{}generated distorted phases listed as Nos.~$18$--$22$, in
Table~III of Hatch et al.\cite{hatch1989ptp}, confirms that our result
is consistent with theirs, though theirs is limited to the $n = 1$ RP
phase.  We confine ourselves to discussion of these five unique
configurations for remainder of this work.

\subsection{RP phases}
\label{subsec:classificationRP}

The RP phases, which we study in this work, consist of periodic arrays
of the above configurations of stacking faults.  While constructing
such arrays from individual stacking faults, the array periodicity may
be maintained either through simple alternating patterns of bulk
regions ($\mathfrak{A/B/A/B}$ sequencing) or through more complex, and
possibly lower\hyp{}energy, patterns (for example,
$\mathfrak{A/B/C/A/B/C}$ sequencing).

Immediately below, we demonstrate that, in fact, an
$\mathfrak{A/B/A/B}$ sequence of bulk regions always corresponds to
repetition of the same type of stacking\hyp{}fault configuration.
Three very compelling reasons then follow for studying RP phases with
this type of periodicity.  First, one should expect the preferred RP
phase to consist of a sequence of stacking faults, which are
\emph{all} of the lowest\hyp{}energy configuration; this would then
naturally correspond to an $\mathfrak{A/B/A/B}$ sequence.  Second,
actual experiments find superlattice sizes consistent with small,
simple repeat units\cite{tilley1977aem, mccarthy1969pei,
mccarthy1969tse}.  Finally, isolated stacking faults are also observed
under certain conditions\cite{tilley1977aem}, and the study of RP
phases of $\mathfrak{A/B/A/B}$ periodicity, containing two identical
faults in each primitive unit cell, then allows for the extraction of
the behavior of individual isolated faults.

To establish that an $\mathfrak{A/B/A/B}$ sequence of bulk regions
corresponds to repetition of the same configuration of stacking fault,
we must prove the equivalence of the two stacking faults,
$(\mathfrak{A/B})$ and $(\mathfrak{B/A})^{\mspace{1mu}\prime}$, in
such an RP phase.  As in
Section~\ref{subsec:classificationStackingFaults}, we take the
algebraic specifier for this first stacking fault $(\mathfrak{A/B})$
to be the generic $(abc \Longrightarrow \alpha\beta\gamma)$ where
$abc$ and $\alpha\beta\gamma$ refer to the rotation states of the two
reference octahedra for this fault, $O_\mathfrak{A}$ and
$O_\mathfrak{B}$ respectively.  We must then choose two corresponding
reference octahedra $O_{\mathfrak{A}}^{\,\prime}$ and
$O_{\mathfrak{B}}^{\,\prime}$ for the second stacking fault
$(\mathfrak{B/A})^{\mspace{1mu}\prime}$.  Since $\mathbf{D} \equiv
-\mathbf{D}$, we can set $O_{\mathfrak{A}}^{\,\prime} = O_\mathfrak{A}
+ (n-1) \zeta a_0 \hat{e}_3$ and $O_{\mathfrak{B}}^{\,\prime} =
O_\mathfrak{B} + (n-1) \zeta a_0 \hat{e}_3$, where $n$ is the number
of the RP phase (A$_{n+1}$B$_n$O$_{3n+1}$) so that $n-1$ layers of
material separate $O_{\mathfrak{A}}$ from
$O_{\mathfrak{A}}^{\,\prime}$ and similarly separate
$O_{\mathfrak{B}}$ from $O_{\mathfrak{B}}^{\,\prime}$.  (Consideration
of the $n = 1$ case should make this apparent.)  The alternating
octahedral rotations of this $a^-a^-a^-$ Glazer system thereby define
the algebraic specifier for the
$(\mathfrak{B/A})^{\mspace{1mu}\prime}$ stacking fault as $
(\,(-1)^{n-1}(\alpha\beta\gamma) \Longrightarrow (-1)^{n-1}(abc)\,)$.
However, displacement of the arbitrary origin (reference octahedra) by
$a_0 [100]$, for cases when $n$ is even, generates an equivalent
algebraic specifier for this $(\mathfrak{B/A})^{\mspace{1mu}\prime}$
interface which is now insensitive to $n$, $(\alpha\beta\gamma
\Longrightarrow abc)$.  Finally, we apply inversion symmetry, as
introduced in Table~\ref{tab:symOps}, to this
$(\mathfrak{B/A})^{\mspace{1mu}\prime}$ stacking fault, $I \circ
(\alpha\beta\gamma \Longrightarrow abc) = (abc \Longrightarrow
\alpha\beta\gamma)$, and thus prove that the $(\mathfrak{A/B})$
stacking fault is equivalent to the
$(\mathfrak{B/A})^{\mspace{1mu}\prime}$ fault.

Lastly, since all stacking faults in an $\mathfrak{A/B/A/B}$ RP phase
always possess equivalent configurations, we can uniquely label these
RP phases using the same rotation\hyp{}sign specifiers established
above for the five symmetry\hyp{}distinct stacking faults: $[0^+]$,
$[0^-]$, $[\frac{\pi}{2}^+]$, $[\pi^+]$, and $[\pi^-]$.

\section{Model and computational methods}
\label{sec:methods}

\subsection{Shell potential}
\label{subsec:methodsShellPot}

As discussed in Section~\ref{sec:background} above, to study the
generic behavior of $a^-a^-a^-$ perovskites which form RP phases, we
employ a shell\hyp{}potential model\cite{dick1958tdc} parameterized
for strontium titanate\cite{akhtar1995css}.  Shell\hyp{}potential
models are formulated as an extension to ionic pair potentials and
employed to capture the polarizability of the atomic constituents.
The shell model separates each ion into two parts, a core and an outer
shell, which possess individual charges that sum to the nominal charge
of the ion.  The total model potential $U$ consists of three terms,
\begin{equation} \label{eqn:u}
  U \equiv U_{\mspace{-1mu}P} + U_{C} + U_{\mspace{-1mu}B},
\end{equation}
representing, respectively, the polarizability of the ions, and the
Coulomb and short\hyp{}range interactions among the ions.  The
polarizability is captured by harmonic springs connecting the core and
shell of each ion, so that $U_{\mspace{-1mu}P}$ has the form,
\begin{equation} \label{eqn:uP}
  U_{\mspace{-1mu}P} = \tfrac{1}{2} \sum_{i} \mspace{1mu} k_i \left|\Delta \mspace{1mu} r_i \right|^2,
\end{equation}
where $|\Delta \mspace{1mu} r_i|$ is the core\hyp{}shell separation
for ion $i$ and the $k_i$ are a set of ion\hyp{}specific spring
constants.  The Coulomb contributions take the form,
\begin{equation} \label{eqn:uC}
  U_{C} = \tfrac{1}{2} \sideset{}{'}\sum_{i,j} \frac{k_c q_i q_j}{r_{ij}},
\end{equation}
where $i$ and $j$ range over all cores and shells (excluding terms
where $i$ and $j$ refer to the same ion), $q_i$ and $q_j$ are the
corresponding charges, $r_{ij}$ is the distance between the charge
centers, and $k_c$ is Coulomb's constant.  Finally, the
short\hyp{}range interactions are included through a sum of
Buckingham\cite{buckingham1958tri} pair potentials (which can be
viewed as combinations of Born\hyp{}Mayer\cite{born1932gi} and
Lennard\hyp{}Jones\cite{lennardjones1931c} potentials) of the form,
\begin{equation} \label{eqn:uB}
  U_{\mspace{-1mu}B} = \tfrac{1}{2} \sum_{i,j} \left( A_{ij} \mspace{1mu} e^{-r_{ij} / \rho_{ij}} - C_{ij} \mspace{1mu} r_{ij}^{-6} \right),
\end{equation}
where $i$ and $j$ range over \emph{all shells} and $A_{ij}$,
$\rho_{ij}$, and $C_{ij}$ are pair\hyp{}specific adjustable
parameters.  Here, the first term (Born\hyp{}Mayer) serves as a
repulsive short\hyp{}range interaction to respect the Pauli exclusion
principle, and the second term (Lennard\hyp{}Jones) models the
dispersion or van der Waals interactions\cite{vanderwaals1873odc}.
The specific electrostatic and short\hyp{}range shell\hyp{}model
parameters used in this study were fit to strontium titanate by Akhtar
et al.\cite{akhtar1995css}, with values as listed in
Tables~\ref{tab:potES} and~\ref{tab:potSR}.  Finally, we wish to
emphasize again, as it is rarely mentioned explicitly in the
shell\hyp{}potential literature, that the pair\hyp{}potential terms in
$U_{\mspace{-1mu}B}$ apply to the \emph{shells only}, and \emph{not}
to the cores.

Shell models have been extensively used for decades as the primary
empirical potential for modeling perovskites and other
oxides\cite{lewis1985pmf, catlow1983iis}.  We tested the correctness
of our coded implementation of this potential through comparisons of
lattice constants and elastic moduli without the AFD reconstruction
and find excellent agreement.  For instance, using the same shell
potential and the same non\hyp{}reconstructed ground state, we predict
a volume per SrTiO$_3$ chemical unit of $59.18$~\AA$^3$, which is
within $0.4$\% of the value calculated by Akhtar et
al.\cite{akhtar1995css} For the elastic moduli, we find $C_{11} =
306.9$~GPa, $C_{12} = 138.7$~GPa, and $C_{44} = 138.8$~GPa, which are
within $1.8$\%, $1.0$\%, and $0.7$\%, respectively, of the values from
Akhtar et al.\cite{akhtar1995css} From this, we conclude that our
implementation of the potential is correct.  We also would like to
note that, when the $a^-a^-a^-$ reconstruction is considered,
significant changes occur, and we find, instead, a volume per
SrTiO$_3$ chemical unit of $58.49$~\AA$^3$ and elastic moduli of
$C_{11} = 275.6$~GPa, $C_{12} = 144.3$~GPa, and $C_{44} = 133.5$~GPa.
This underscores the importance of considering AFD reconstructions
when constructing such potentials.

\begin{table}
  \setlength{\doublerulesep}{0\doublerulesep}
  \setlength{\tabcolsep}{4.9\tabcolsep} 
  \begin{tabular}{cD{.}{.}{3.3}D{.}{.}{2.3}D{.}{.}{5.3}}
    \hline\hline\\[-1.5ex]
    \multirow{2}{*}{Ion} & \multicolumn{1}{c}{Shell} & \multicolumn{1}{c}{Core} & \multicolumn{1}{c}{Spring Constant}\\
    & \multicolumn{1}{c}{Charge [e]} & \multicolumn{1}{c}{Charge [e]} & \multicolumn{1}{c}{[eV$\cdot$\AA$^{-2}$]}\\[0.5ex]
    \hline\\[-1.5ex]
    Sr$^{2+}$ &   1.526 &  0.474 &    11.406\\
    Ti$^{4+}$ & -35.863 & 39.863 & 65974.0\\
    O$^{2-}$  &  -2.389 &  0.389 &    18.41\\[0.5ex]
    \hline\hline
  \end{tabular}
  \caption[Electrostatic shell\hyp{}model potential parameters used in
    this study]{Electrostatic shell\hyp{}model potential parameters
    used in this study (from Akhtar et al.\cite{akhtar1995css}).}
  \label{tab:potES}
\end{table}
\begin{table}
  \setlength{\doublerulesep}{0\doublerulesep}
  \setlength{\tabcolsep}{5\tabcolsep}
  \begin{tabular}{rD{.}{.}{5.2}D{.}{.}{1.5}D{.}{.}{2.1}}
    \hline\hline\\[-1.5ex]
    \multicolumn{1}{c}{Interaction} & \multicolumn{1}{c}{A [eV]} & \multicolumn{1}{c}{$\rho$ [\AA]} & \multicolumn{1}{c}{C [eV$\cdot$\AA$^6$]}\\[0.5ex]
    \hline\\[-1.5ex]
    Sr$^{2+}$ $\Leftrightarrow$ O$^{2-}$ &   776.84 & 0.35867 &  0.0\\
    Ti$^{4+}$ $\Leftrightarrow$ O$^{2-}$ &   877.20 & 0.38096 &  9.0\\
    O$^{2-}$ $\Leftrightarrow$ O$^{2-}$  & 22764.3  & 0.1490  & 43.0\\[0.5ex]
    \hline\hline
  \end{tabular}
  \caption[Short\hyp{}range shell\hyp{}model potential parameters used
    in this study]{Short\hyp{}range shell\hyp{}model potential
    parameters used in this study (from Akhtar et
    al.\cite{akhtar1995css}).}
  \label{tab:potSR}
\end{table}

\subsection{Numerical methods}
\label{subsec:methodsNumerical}

In this work, we compute the Coulombic interaction\cite{ewald1921boe}
from \eqref{eqn:uC} using a Particle Mesh Ewald
algorithm\cite{darden1993pme, essmann1995asp, deserno1998htm} with all
real\hyp{}space pair\hyp{}potential terms computed out to a fixed
cutoff distance using neighbor tables.  Analytic derivatives are used
to determine the forces on the cores and shells within the supercell,
and finite differences are used to compute the generalized forces on
the superlattice vectors.  While such finite\hyp{}difference methods
occasionally introduce issues with numerical precision into the
resulting calculations, we are able to mitigate such effects with care
in selection of the finite\hyp{}difference step, scaling it
proportionally to the sizes of the lattice vectors involved.

With the gradients determined as above, we relax each system using the
technique of preconditioned conjugate gradient
minimization\cite{hestenes1952mcg} (specifically, the
Polak\hyp{}Ribi\`{e}re\cite{polak1969ncm} method), fully optimizing
the ionic coordinates in an inner loop of the routine and then
optimizing the lattice vectors in an outer loop.  The preconditioner
applies only to the ionic relaxation of this procedure and scales the
generalized force on the geometric center of each ion (mean position
of the core and shell) separately from the generalized force on the
core\hyp{}shell displacement, with the latter scaled in inverse
proportion to the spring constant of the corresponding ion.  This
process ensures that all supercell energies are relaxed with respect
to both ionic and lattice coordinates to a precision of
$\sim$$0.3$~\micro{}eV per supercell.

\subsection{Ground state}
\label{subsec:methodsGroundState}

To establish the ground state of the model potential used in our
calculations, we have carried out what we regard as a thorough, but
not exhaustive, search for a \emph{probable} ground\hyp{}state
structure.  Indeed, we have found no alternative structure which
relaxes to an energy less than our candidate ground\hyp{}state
structure within our potential.  We performed quenches on hundreds of
random displacements from the idealized positions of the $1 \times 1
\times 1$ primitive unit cell to explore various potential
reconstructions for supercells up to $6\times 6 \times 6$.  We also
considered a number of highly ordered configurations commensurate with
antiferrodistortive disordering.

Among those minima which we explored, we selected the
lowest\hyp{}energy configuration to serve as the bulk crystalline
state throughout this study.  This configuration possesses a fairly
regular pattern as depicted in Figure~\ref{fig:reconstruction}, namely
each oxygen octahedron rotates slightly along trigonal directions in a
cell\hyp{}doubling, alternating three\hyp{}dimensional $2 \times 2
\times 2$ checkerboard pattern, which is precisely the desired
$a^-a^-a^-$ Glazer system from Section~\ref{sec:background}.  We also
find that the lattice vectors assume the rhombohedral symmetry commonly
found in $a^-a^-a^-$ perovskites, with the cubic lattice stretching
along the same trigonal axes about which the octahedra rotate.

The specific lattice vectors which we find for the $2 \times 2 \times
2$ reconstructed supercell of our model are the columns of the
following matrix,
\begin{equation*} \label{eqn:vecSC}
  \mathbf{R} = \left(
  \setlength{\arraycolsep}{2\arraycolsep}
  \begin{array}{D{.}{.}{1.3}D{.}{.}{1.3}D{.}{.}{1.3}}
     7.764 &  0.028 &  0.028\\
     0.028 &  7.764 &  0.028\\
     0.028 &  0.028 &  7.764\\
  \end{array}
  \right) \text{\AA}.
\end{equation*}

\subsection{Construction of RP supercells}
\label{subsec:methodsConstruction}

Our construction of initial ionic configurations for the
A$_{n+1}$B$_n$O$_{3n+1}$ RP phases proceeds through a detailed process
defined in Section~\ref{sec:classification}.  As discussed there, we
carefully consider a large number of combinatorial possibilities
resulting from the different geometries due to the aforementioned $2
\times 2 \times 2$ cell\hyp{}doubling, $a^-a^-a^-$ bulk
reconstruction.  We construct each configuration of the RP phase by
stacking $2n$ layers of bulk perovskite along the $[001]$ direction,
with the two stacking faults (excess AO $(001)$ planes) so situated as
to equally divide the supercell into two bulk slabs, each with $n$
layers and potentially different $a^-a^-a^-$ reconstructions.
Relative to each other, these bulk slabs are also laterally displaced
by one of the four (equivalent) $\frac{a_0}{2}[\pm1,\pm1,0\,]$ vectors
at each stacking fault.  We perform this procedure to create all eight
$(111 \Longrightarrow \alpha\beta\gamma)$ configurations from
Section~\ref{subsec:classificationStackingFaults}, through
combinations of reconstructions in the two bulk slabs, for RP phases
from $n = 1\ldots30$, containing from $56$ to $1216$ ions,
respectively.  Each RP superlattice system is then fully relaxed, as
described above, both in terms of internal ionic coordinates and
lattice vectors, until the energy is minimized to within a precision
of $\sim$$0.3$~\micro{}eV\@.  In all cases, we find that the material
on either side of the stacking fault maintains a uniform bulk
reconstruction throughout, so that the enumeration of
Section~\ref{subsec:classificationStackingFaults} is suitably
complete.  Finally, we test for meta\hyp{}stability by creating
ensembles of $20$--$250$ samples for each configuration of octahedra
rotations for all RP phases up to $n = 10$, introducing
root\hyp{}mean\hyp{}square displacements of $0.001$~\AA\ for each
coordinate axis of every core and shell, and quenching each of these
randomized systems.

\section{Results}
\label{sec:results}

As described in Section~\ref{sec:background} above, the results
obtained below have been computed with the standard
shell\hyp{}potential model for strontium titanate\cite{akhtar1995css,
lewis1985pmf, catlow1983iis}, but are interpreted to represent the
generic behavior of an antiferrodistortive perovskite (ABO$_3$) from
among the class of perovskites within the $a^-a^-a^-$ Glazer system
which form RP phases.

\subsection{Formation of RP phases}
\label{subsec:resultsFormation}

After constructing the various possible RP phases according to the
prescription in Section~\ref{subsec:methodsConstruction}, the ideal procedure for
determining the energetics of the RP phases is to fully relax each
system, in terms of \emph{both} internal atomic coordinates \emph{and}
superlattice vectors, without any externally imposed symmetries.  The
energies of the reference bulk materials, perovskite (ABO$_3$) and
A\hyp{}oxide (AO), should also be fully relaxed in the same sense.
 
Without full relaxation of lattice vectors for both the RP phases and
bulk reference cells, any attempts to extract energies will introduce
significant errors that scale in proportion to the amount of bulk
material between stacking faults.  (Such failures to fully relax the
lattice vectors might arise due to expediency in \emph{ab initio}
studies or through attempts to simulate higher\hyp{}temperature phases
by imposing a higher\hyp{}temperature lattice structure while relaxing
ionic coordinates.)

\begin{figure}
  \begin{centering}
    \subfloat[][]{\includegraphics[width=\columnwidth]{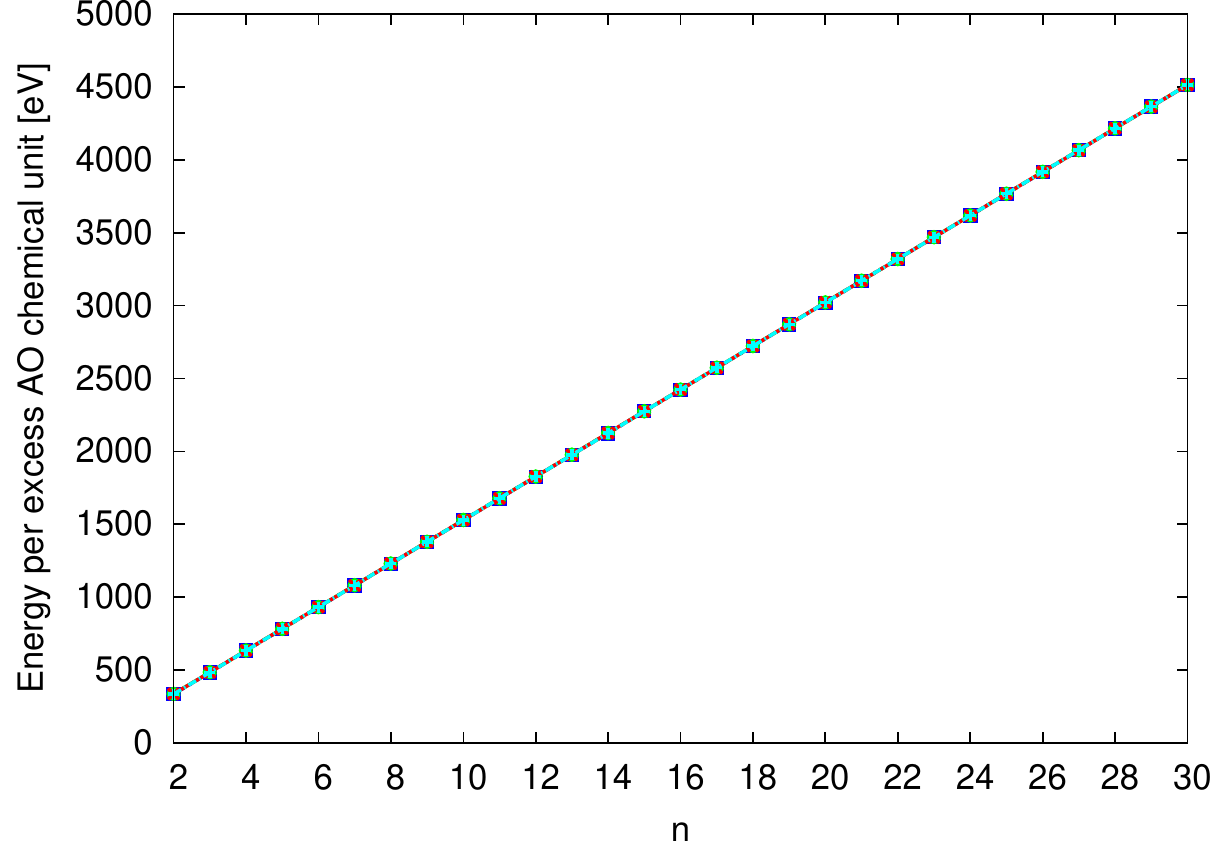}}\\
    \subfloat[][]{\includegraphics[width=\columnwidth]{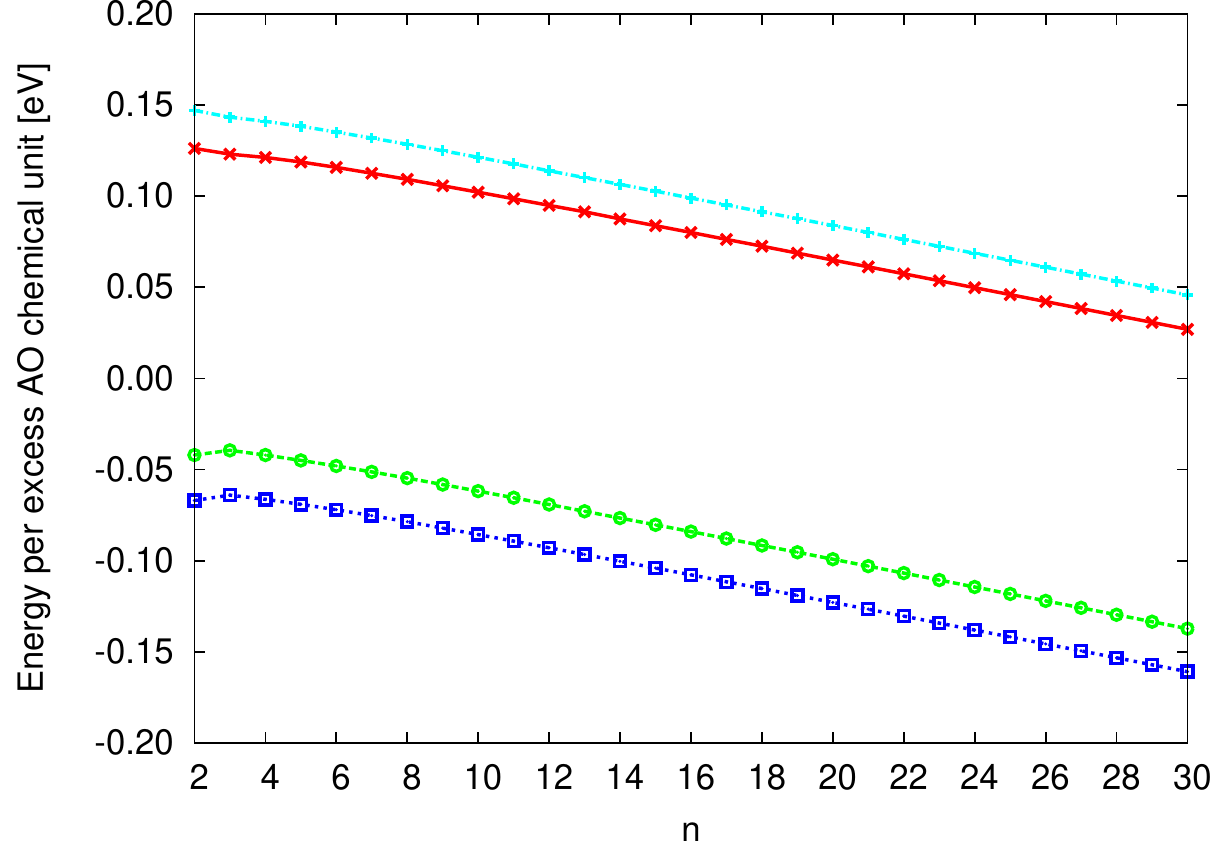}}
    \caption[Formation energies of RP phases with incorrect
      supercells]{Formation energies of RP phases per excess AO
      chemical unit as a function of $n$ for various members of
      homologous series A$_{n+1}$B$_n$O$_{3n+1}$: methodological error
      of employing (a) tetragonal supercells for RP phases or (b)
      fully relaxed supercells for RP phases but unrelaxed lattice
      vectors for bulk reference cells.  Each panel shows results for
      the distinct stacking\hyp{}fault configurations corresponding to
      different arrangements of octahedral rotations, as enumerated in
      the text (Section~\ref{subsec:classificationStackingFaults}).}
    \label{fig:formEnergiesFail}
  \end{centering}
\end{figure}
Figure~\ref{fig:formEnergiesFail}(a) shows actual results obtained
from such a methodologically flawed procedure, where the energy needed
to form the RP phase from the bulk perovskite ABO$_3$ and bulk
A\hyp{}oxide AO, $E_{\text{A}_{n+1}\text{B}_n\text{O}_{3n+1}} - n
\cdot E_{\text{ABO}_3} - E_{\text{AO}}$, is computed without full
relaxation of the lattice vectors of the RP phase, but instead with
the enforcement of tetragonal lattice vectors associated with a
higher\hyp{}temperature symmetry not possessed by the ground state of
our model.  Under these constraints, the energy exhibits a clear
linear \emph{increase} with $n$, the amount of bulk material between
stacking faults, due to an inflated RP\hyp{}phase energy associated
with artificial strain in the bulk regions.  (Note that the results in
the figure are visually indistinguishable whether or not one allows
the perovskite or A\hyp{}oxide bulk reference cells to fully relax.)
Figure~\ref{fig:formEnergiesFail}(b) shows results obtained where the
lattice vectors of the RP phases are allowed to relax fully but the
lattice vectors of the bulk reference cells are not.  In this case,
excess bulk energy is not included in the RP cell but instead in the
bulk reference cells, and the energy of formation now \emph{decreases}
linearly with $n$.  Either type of error will confound the extraction
of meaningful information on the RP phases.  The calculation of such
phases is herein seen to be quite delicate, a fact which only becomes
obvious when calculations are pursued for large values of $n$.

\begin{figure}
  \begin{centering}
    \includegraphics[width=\columnwidth]{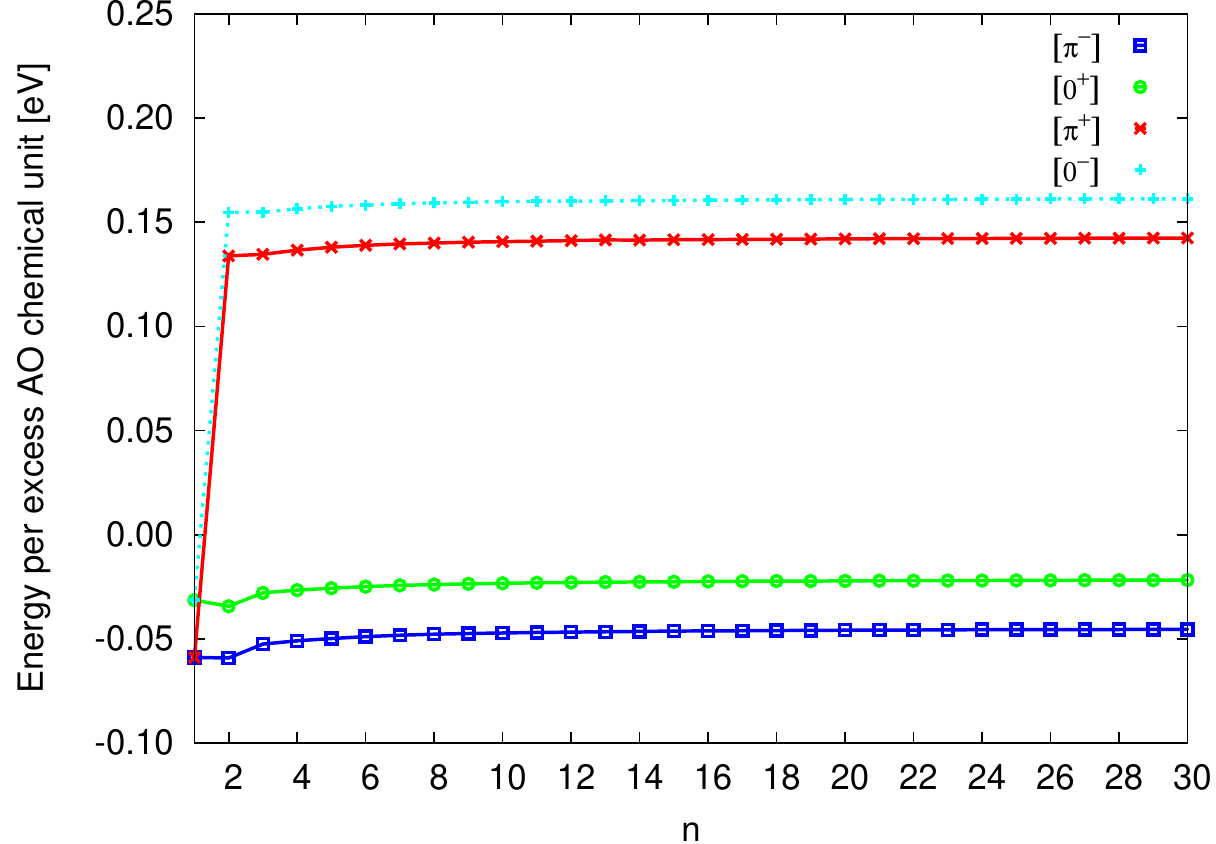}
    \caption[Formation energies of RP phases]{Formation energies of RP
      phases per excess AO chemical unit as a function of $n$ for
      various members of homologous series A$_{n+1}$B$_n$O$_{3n+1}$:
      labeled curves represent each of the stable stacking\hyp{}fault
      configurations corresponding to different arrangements of
      octahedral rotations, as enumerated in the text
      (Section~\ref{subsec:classificationStackingFaults}).  Results
      calculated using fully relaxed ionic \emph{and} lattice
      coordinates for both RP phases and bulk references.}
    \label{fig:formEnergiesRP}
  \end{centering}
\end{figure}

Figure~\ref{fig:formEnergiesRP} presents the results, properly
calculated with full relaxation of all lattice vectors, for the
energies of all eight $(111 \Longrightarrow \alpha\beta\gamma)$
stacking\hyp{}fault configurations of our model potential, relative to
the corresponding amounts of the fully relaxed, reconstructed
perovskite and the fully relaxed A\hyp{}oxide material.  As in the
examples above, the energies are expressed per excess AO chemical
unit, $2 \times 2 = 4$ of which appear in each reconstructed RP
superlattice unit cell (and $2 \times (2 \times 2) = 8$ of which
appear in each supercell with two stacking faults and thus two RP
superlattice unit cells).  The final relaxed RP configurations indeed
are not distinct, as expected from the symmetry arguments in
Section~\ref{subsec:classificationStackingFaults}, which showed that
$[\frac{\pi}{2}^+] \equiv [\frac{\pi}{2}^-] \equiv
[\frac{\smash{3}\pi}{2}^+] \equiv [\frac{\smash{3}\pi}{2}^-]$, so that
there should be at most five different configurations.  Somewhat
unexpectedly, rather than displaying five separate curves for the five
symmetry\hyp{}distinct configurations ($[0^+]$, $[0^-]$,
$[\frac{\pi}{2}^+]$, $[\pi^+]$, $[\pi^-]$),
Figure~\ref{fig:formEnergiesRP} instead shows only four curves: we
find that the $[\frac{\pi}{2}^+]$ configuration is unstable and
relaxes directly to the $[\pi^-]$ configuration, as determined by
energy calculations and observations of transitions in lattice
symmetries for each configuration during full relaxation.
Additionally, the $[0^-]$ configuration appears meta\hyp{}stable,
often transitioning, upon small perturbations described in
Section~\ref{subsec:methodsConstruction}, to the $[0^+]$
configuration.  The remainder of this discussion thus only reports on
the behavior of the final three distinct, stable configurations
($[0^+]$, $[\pi^+]$, $[\pi^-]$) as well as the one additional
distinct, meta\hyp{}stable configuration ($[0^-]$).

From Figure~\ref{fig:formEnergiesRP}, we note immediately that,
depending upon the relative rotational orientation of oxygen octahedra
on either side of the stacking faults, the configurations separate
into two groupings, one with positive energy and one with negative
energy relative to phase separation into bulk perovskite and bulk
A\hyp{}oxide.  The energy scales associated with the different
configurations may thus be sufficient to either stabilize or
destabilize the formation of an RP phase.

Moreover, for our model potential, the two more stable
stacking\hyp{}fault configurations ($[0^+]$ and $[\pi^-]$) possess
negative energies for all $n$, providing the first demonstration that
RP phases can be favored over formation of bulk perovskite and
A\hyp{}oxide for \emph{all} values of $n$.  Also, we find the
$[\pi^-]$ configuration to be favored over all other possible
configurations in our model, regardless of the value of $n$.

It is also immediately apparent that the RP phases for $n = 1$ deserve
special consideration for our model potential.  The $n = 1$ phases for
the two configurations with positive energy collapse to related $n =
1$ configurations with negative energy, with $[\pi^+]$ becoming
$[\pi^-]$ and $[0^-]$ becoming $[0^+]$.  This might be related to the
realization that the full $2 \times 2 \times 2$ reconstruction cannot
be expressed on either side of the stacking fault for an $n = 1$ RP
phase, and thus these higher\hyp{}energy configurations forfeit some
stability for the $n = 1$ RP phase.

The magnitude of the energy differences between configurations is not
only sufficient to affect the stability of the phases, but is also
significant compared to thermal energy scales.  For instance, we find
the separation between the stable and unstable groupings to be
$\sim$$200$~meV per excess AO chemical unit (or $\sim$$800$~meV per RP
superlattice unit cell, $(2 \times 2) \cdot
$A$_{n+1}$B$_n$O$_{3n+1}$).  To place this energy in context, we can
also consider the energy per interfacial oxygen octahedron (one on the
top and bottom of the stacking fault for each excess AO chemical
unit), $\sim$$100$~meV, which is about $\sim$$30$~meV per
configurational (rotational) degree of freedom.  This is larger than,
but comparable to, the thermal energy at room temperature, $26$~meV,
so that at significant temperatures we still can expect a small, but
statistically noticeable, tendency for the stacking fault to
preferentially select the minimal energy configuration over other
disfavored configurations.

This energy scale is also significantly larger than the energy scales
we might expect for excitations in the bulk of the material.  For
instance, the structural phase transition in strontium titanate (the
material to which the original model actually had been
fit\cite{akhtar1995css}) occurs near
$105$~K\cite{vonwaldkirch1973fsn}, which corresponds to an energy
scale of $9$~meV\@.  To further place the above $\sim$$30$~meV
interfacial energy scale in context, we consider the energies within
our model of two alternate rotational $2 \times 2 \times 2$ bulk
reconstructions, those with rotations about $\langle 110 \rangle$ axes
(orthorhombic) and $\langle 100 \rangle$ axes (tetragonal), finding
rotational energy scales of $4$~meV and $16$~meV, respectively,
relative to the ground\hyp{}state reconstruction.  These comparisons
lead to the intriguing conjecture that, for some perovskites with low
densities of stacking faults (high $n$), there exist a range of
temperatures above the bulk structural phase transitions, in which the
bulk rotational energy scale is insufficient to constrain the
rotations of the octahedra, while the interfacial energy scale is
suitable to constrain the octahedral rotations to specific preferred
orientations.  This corresponds to a picture of ``fuzzy'' octahedra in
the bulk regions with noticeable orientational preferences remaining
for the octahedra at the stacking faults.  Alternatively, for
perovskites in RP phases with low values of $n$ and higher densities
of stacking faults, one might expect this increased energy scale to
raise the transition temperature associated with octahedral rotations.

\subsection{Stacking fault interactions and\texorpdfstring{\\}{ }intergrowth formation}
\label{subsec:resultsInteraction}

\begin{figure}
  \begin{centering}
    \includegraphics[width=\columnwidth]{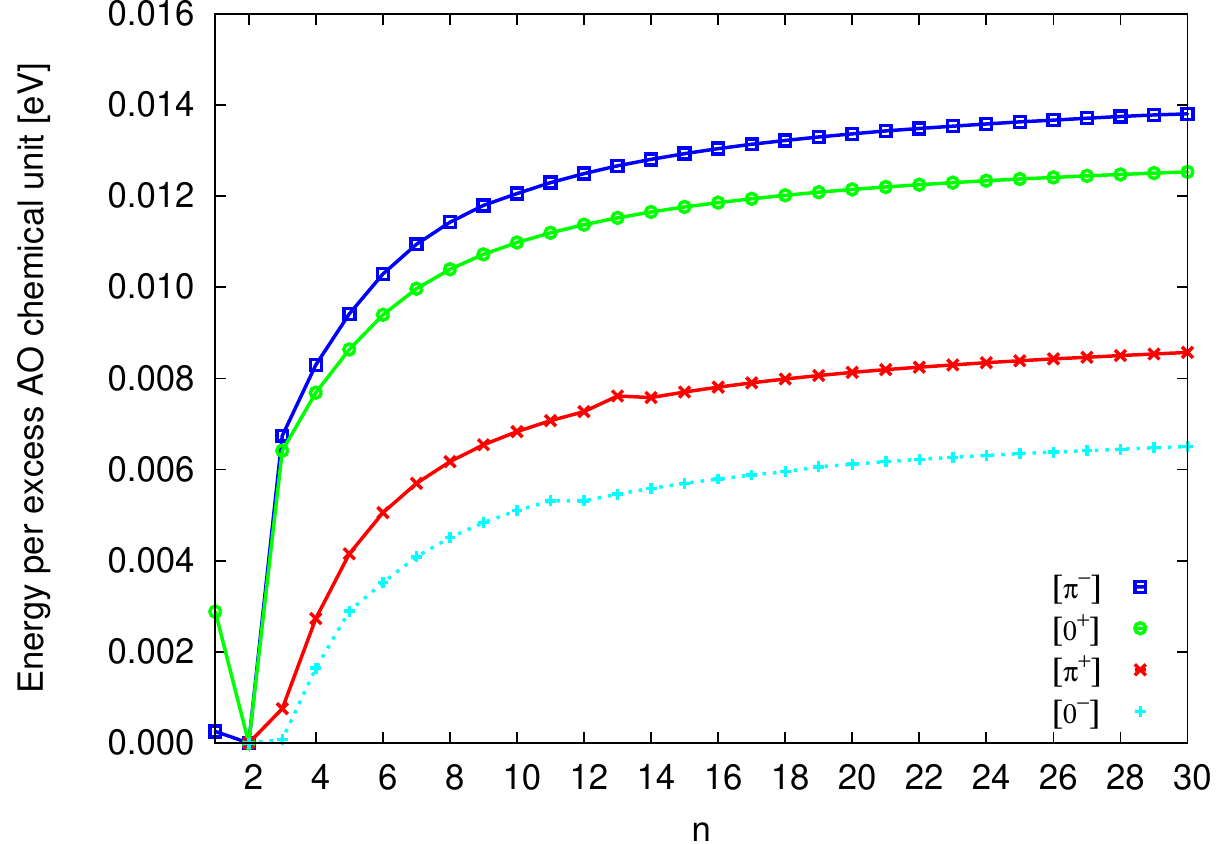}
    \caption[Interaction energies among stacking faults]{Interaction
      energies among stacking faults (formation energies relative to
      $n = 2$ RP phase) per excess AO chemical unit as a function of
      $n$ for the homologous series A$_{n+1}$B$_n$O$_{3n+1}$ (same
      notations as in Figure~\ref{fig:formEnergiesRP}).  Data depict
      relatively weak, but consistently attractive, interactions in
      all cases.}
    \label{fig:interEnergiesRP}
  \end{centering}
\end{figure}

While Section~\ref{subsec:resultsFormation} discussed the relationship
among energies of formation of different configurations, it is further
instructive to examine the energy and behavior of each configuration
independently, as a function of separation between stacking faults.
Figure~\ref{fig:interEnergiesRP} plots the interaction energies per
excess AO chemical unit for each configuration, which, for a given
configuration, are the formation energies of each phase
\emph{relative} to the energy of its $n = 2$ phase. (For the two
stable, low\hyp{}energy configurations within our model, the $n = 2$
phase is, in fact, lowest in energy, so that this plot represents the
binding energies of the stacking faults for these configurations.  For
the two high\hyp{}energy configurations, unstable to dissociation into
separate regimes of bulk perovskite and A\hyp{}oxide, the $n = 1$
phase is actually lowest in energy, so that this plot approximately
represents the binding energies of the stacking faults.)  For our
model, in all cases, there is a relatively weak, but consistently
attractive, interaction (apart from special cases for $n = 1$),
varying by over a factor of two from $\sim$$7$--$14$~meV per excess AO
chemical unit, depending upon the stacking\hyp{}fault configuration.
The strengths of the interaction occur in precisely the same sequence
as the stability of the configurations, with the more attractive
interactions occurring in the more stable configurations.

Within the current model, the universally attractive and eventually
asymptotic nature of the interaction for $n \ge 2$ leads to a set of
interaction energies with \emph{downward} curvature.  The standard
secant construction for mixed phases then indicates that, for $n > 2$,
the energy of a system can always be lowered by forming a
so\hyp{}called ``intergrowth'' phase, mixing regions of lower\hyp{}
and higher\hyp{}$n$ phases (i.e., non\hyp{}uniformly changing the
separation between stacking faults that define the phases) while
maintaining a consistent average stoichiometry.  To understand how
this arises from the particular form of these interactions, note that
changing the location of a particular fault (in a uniform $n > 2$
phase) decreases the distance to one of its neighboring faults but
increases the distance to the other neighboring fault.  The downward
curvature of the interaction energy, however, ensures that the energy
gain from the decreased distance more than compensates for the cost
from the increased distance, so that the overall energy is lowered by
varying the distances between faults.

Finally, we consider the possibility of intergrowth phenomena.  The
two higher\hyp{}energy configurations within our model are unstable
with respect to dissociation into intergrowths of bulk perovskite and
bulk A\hyp{}oxide, and thus need not be considered further here.  To
examine the possibility of intergrowths in the two lower\hyp{}energy
configurations, we consider environments both with a relatively low
density (high $n$) of stacking faults and those with a higher density
(lower $n$) of stacking faults.  In the former case, the stacking
faults attract to form dense phases at the value of $n$ for which the
energy is minimized, namely $n = 2$ within our model, alternating with
regions of bulk perovskite.  In the latter case, the additional
A\hyp{}oxide exceeds the concentration required for creation of a
consistent phase at the preferred minimum energy $n = 2$ value.  For
the stoichiometry such that $1 < n < 2$ on \emph{average} throughout
our model system, the excess A\hyp{}oxide could either form
intergrowths consisting of the $n = 2$ phase and bulk A\hyp{}oxide or
intergrowths consisting of $n = 2$ and $n = 1$ phases.  The creation
of $n = 1$ phases, by insertion of excess AO planes, corresponds to
the transformation of a given unit of the $n = 2$ phase into two units
of the $n = 1$ phase.  Since twice the formation energy of the $n = 1$
phase is always lower than the formation energy of the $n = 2$ phase
within our model, we recognize that, for stoichiometries with $1 < n <
2$ on average, the two stable configurations ($[0^+]$ and $[\pi^-]$)
will always form intergrowths of $n = 1$ and $n = 2$ phases, rather
than of bulk A\hyp{}oxide and $n = 2$ phase.  For RP phases with yet
higher concentrations of species A ($n < 1$), a similar argument shows
that, within our model, intergrowths form between the $n = 1$ RP phase
and bulk A\hyp{}oxide.

\subsection{Isolated stacking faults}
\label{subsec:resultsIsolated}

Although the stacking faults are attractive in our particular model,
crucial early growth experiments on RP phases often observed these
stacking faults in isolation\cite{tilley1977aem}, perhaps due to
kinematic constraints.  Here, we consider the physical properties of
such isolated faults.

\begin{figure}
  \begin{centering}
    \includegraphics[width=\columnwidth]{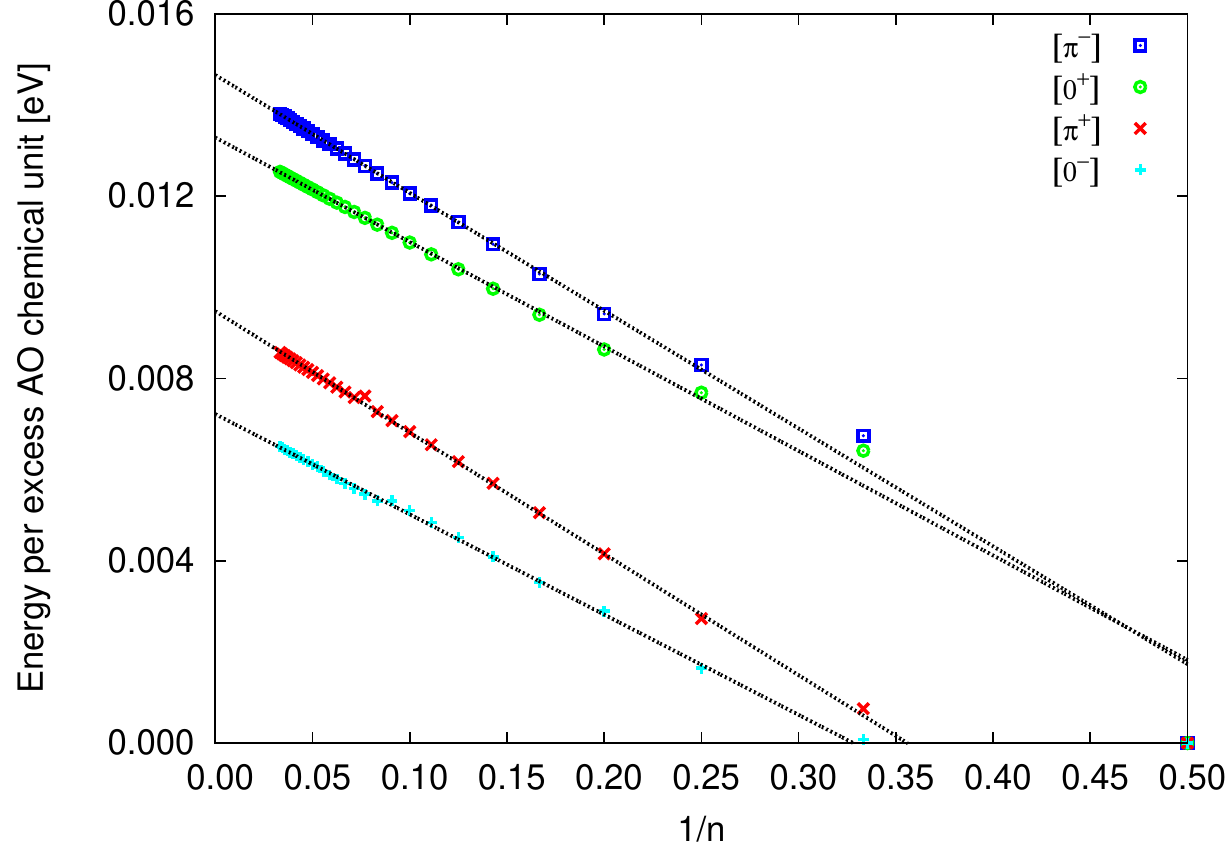}
    \caption[Interaction energies among stacking faults as $n
      \rightarrow \infty$]{Interaction energies among stacking faults
      (formation energies relative to $n = 2$ RP phase) per excess AO
      chemical unit as a function of $1 / n$ for the homologous series
      A$_{n+1}$B$_n$O$_{3n+1}$: extraction of energies to infinite
      superlattice size (dotted lines) to determine binding energies
      of stacking faults.  Linear behavior indicates an
      inverse\hyp{}distance interaction between stacking faults.}
    \label{fig:interEnergiesIsoSF}
  \end{centering}
\end{figure}
\begin{table}
  \setlength{\doublerulesep}{0\doublerulesep}
  \setlength{\tabcolsep}{5\tabcolsep}
  \begin{tabular}{cD{.}{.}{1.3}}
    \hline\hline\\[-1.5ex]
    Configuration & \multicolumn{1}{c}{Binding Energy [meV]}\\[0.5ex]
    \hline\\[-1.5ex]
    $[0^-]^\dagger$             &  7.218\\[0.3ex]
    $[\pi^+]\phantom{^\dagger}$ &  9.476\\[0.3ex]
    $[0^+]\phantom{^\dagger}$   & 13.288\\[0.3ex]
    $[\pi^-]\phantom{^\dagger}$ & 14.658\\[0.5ex]
    \hline\hline\\[-2ex]
    \multicolumn{1}{l}{\footnotesize $^\dagger$ Meta-stable}
  \end{tabular}
  \caption[Binding energies of stacking faults]{Binding energies of
    stacking faults per excess AO chemical unit for different
    configurations.}
  \label{tab:interEnergiesIsoSF}
\end{table}

To extract the behavior of isolated stacking faults at infinite
separation, we first determine the functional dependence of the
interaction on the separation between stacking faults.
Figure~\ref{fig:interEnergiesIsoSF} shows the interaction energies of
the stacking faults from Figure~\ref{fig:interEnergiesRP}, now
plotting these energies as a function of $1 / n$, where $n$ measures
the separation between stacking faults in terms of the number of
intervening bulk layers.  The data in this figure exhibit clear linear
behavior, indicating that the interaction is inversely proportional to
fault separation.  The ordinate intercepts ($1 / n \rightarrow 0$, or
$n \rightarrow \infty$) denote the energy released per excess AO
chemical unit when stacking faults bind together into RP phases at
their most energetically favored value of $n$ (for the two stable
configurations within our model, $[0^+]$ and $[\pi^-]$, with energy
minima at $n = 2$).

Table~\ref{tab:interEnergiesIsoSF} summarizes the values of these
binding energies per excess AO chemical unit, as determined by linear
least\hyp{}square fits to the energies for RP phases in
Figure~\ref{fig:interEnergiesIsoSF}, ignoring in each case the three
smallest separations ($n \le 3$) which clearly involve significant
higher\hyp{}order interactions.  Even for the most attractive
configuration ($[\pi^-]$), our model shows a $\sim$$7$~meV binding
energy per octahedron neighboring the stacking fault (recall that
there are two interfacial octahedra for each excess AO chemical unit,
one on either side of the stacking fault), which is relatively low
compared to room temperature, a fact which may be related to the
observation of isolated faults in some materials.

\begin{figure}
  \begin{centering}
    \includegraphics[width=\columnwidth]{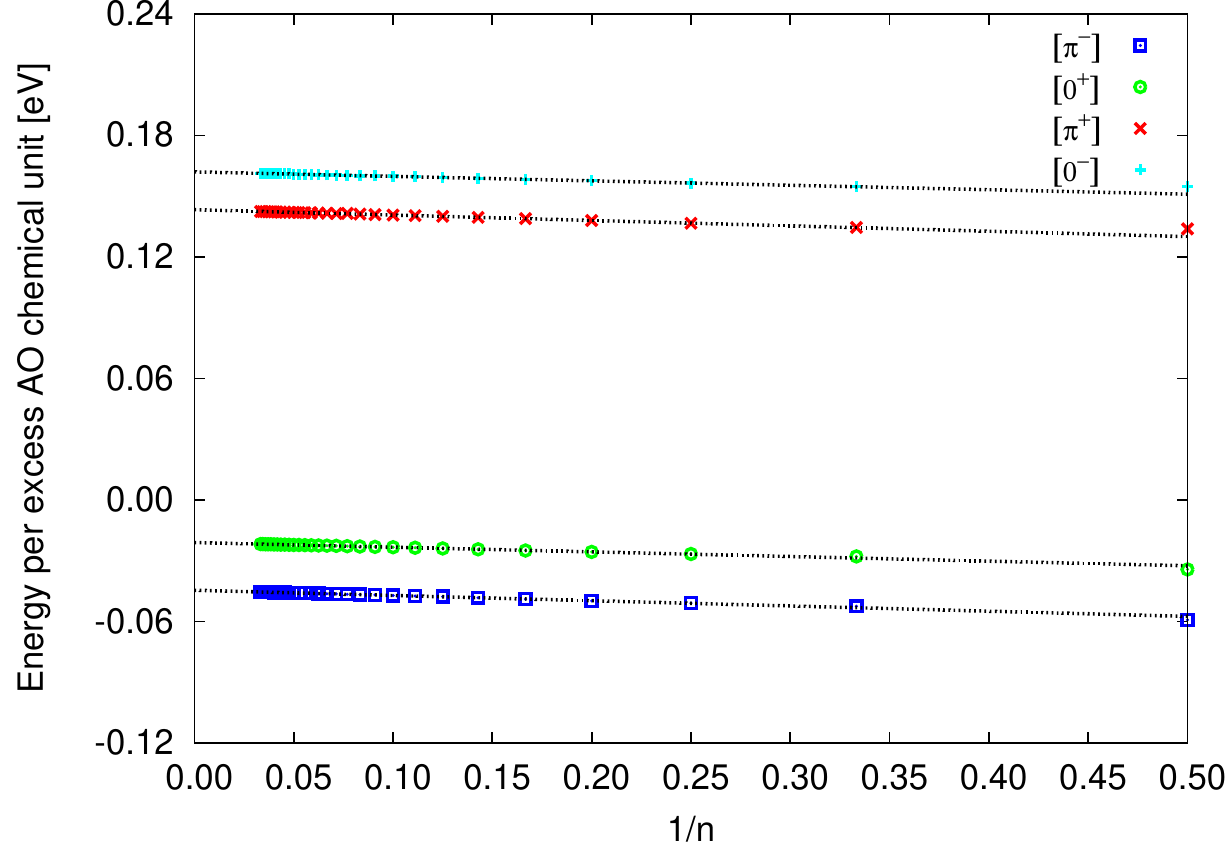}
    \caption[Formation energies of isolated stacking faults]{Formation
      energies of RP phases per excess AO chemical unit as a function
      of $1 / n$ for the homologous series A$_{n+1}$B$_n$O$_{3n+1}$:
      extraction of energies to infinite superlattice size (dotted
      lines) to determine formation energies of isolated stacking
      faults.}
    \label{fig:formEnergiesIsoSF}
  \end{centering}
\end{figure}
\begin{table}
  \setlength{\doublerulesep}{0\doublerulesep}
  \setlength{\tabcolsep}{5\tabcolsep}
  \begin{tabular}{cD{.}{.}{1.3}}
    \hline\hline\\[-1.5ex]
    Configuration & \multicolumn{1}{c}{Formation Energy [meV]}\\[0.5ex]
    \hline\\[-1.5ex]
    $[0^-]^\ddagger$             &  162.0\\[0.3ex]
    $[\pi^+]\phantom{^\ddagger}$ &  143.3\\[0.3ex]
    $[0^+]\phantom{^\ddagger}$   &  -21.0\\[0.3ex]
    $[\pi^-]\phantom{^\ddagger}$ &  -44.5\\[0.5ex]
    \hline\hline\\[-2ex]
    \multicolumn{1}{l}{\footnotesize $^\ddagger$ Meta-stable}
  \end{tabular}
  \caption[Formation energies of isolated stacking faults]{Formation
    energies of isolated stacking faults per excess AO chemical unit
    for different configurations.}
  \label{tab:formEnergiesIsoSF}
\end{table}

With the form of the interaction now determined, we turn to the
formation energies of isolated stacking faults in their various
configurations.  We can now finally determine whether these different
configurations have a large effect on formation energies of RP phases,
as Section~\ref{subsec:resultsFormation} suggests.
Figure~\ref{fig:formEnergiesIsoSF} shows the formation energy per
excess AO chemical unit as a function of $1 / n$.

Here, the ordinate intercepts give the energy to incorporate an AO
chemical unit into an isolated, excess AO plane in perovskite relative
to that chemical unit appearing in bulk A\hyp{}oxide.
Table~\ref{tab:formEnergiesIsoSF} summarizes the formation energies of
isolated stacking faults as determined by the same type of linear
least\hyp{}square fit used to extract the binding energies above.
Within our model, we again find significant differences in formation
energies between the two groupings of configurations (those with
positive and negative energies), of up to $\sim$$200$~meV per excess
AO chemical unit or $\sim$$100$~meV per oxygen octahedron neighboring
the stacking fault.  We also identify a much smaller energy difference
between the two lowest\hyp{}energy configurations (those with negative
energies) of $23.5$~meV per excess AO chemical unit, or $11.8$~meV per
octahedron.  Following a similar argument as in
Section~\ref{subsec:resultsFormation}, for a range of temperatures
above the bulk phase transitions, we may still find that octahedra
near isolated stacking faults are constrained to a subset of
low\hyp{}energy rotational orientations, while those in the bulk
fluctuate among all possible rotations.

\begin{figure}
  \begin{centering}
    \subfloat[][{High-energy configurations: $[0^-]$ or $[\pi^+]$}]{\includegraphics[width=0.8\columnwidth]{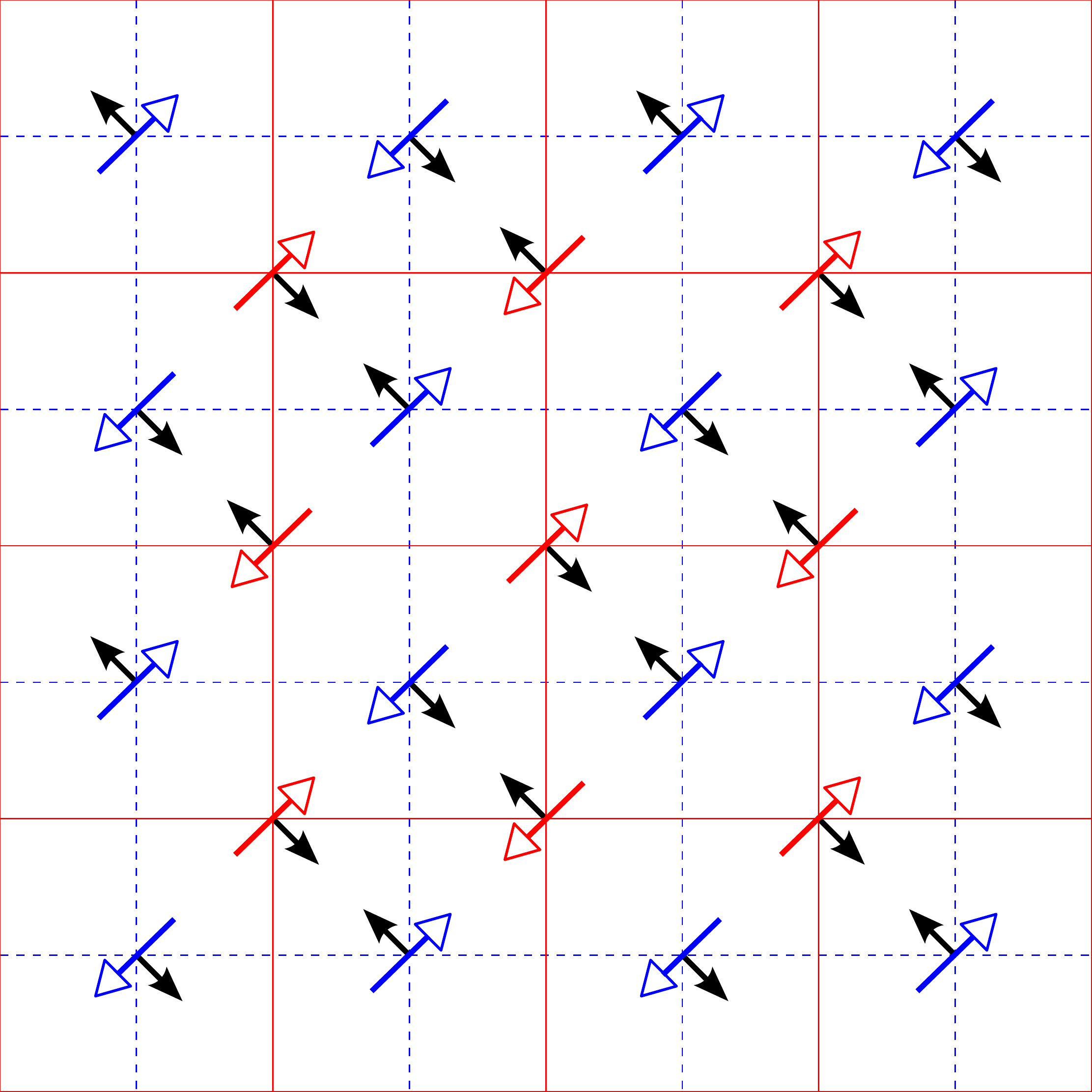}}\\
    \subfloat[][{Low-energy configurations: $[0^+]$ or $[\pi^-]$}]{\includegraphics[width=0.8\columnwidth]{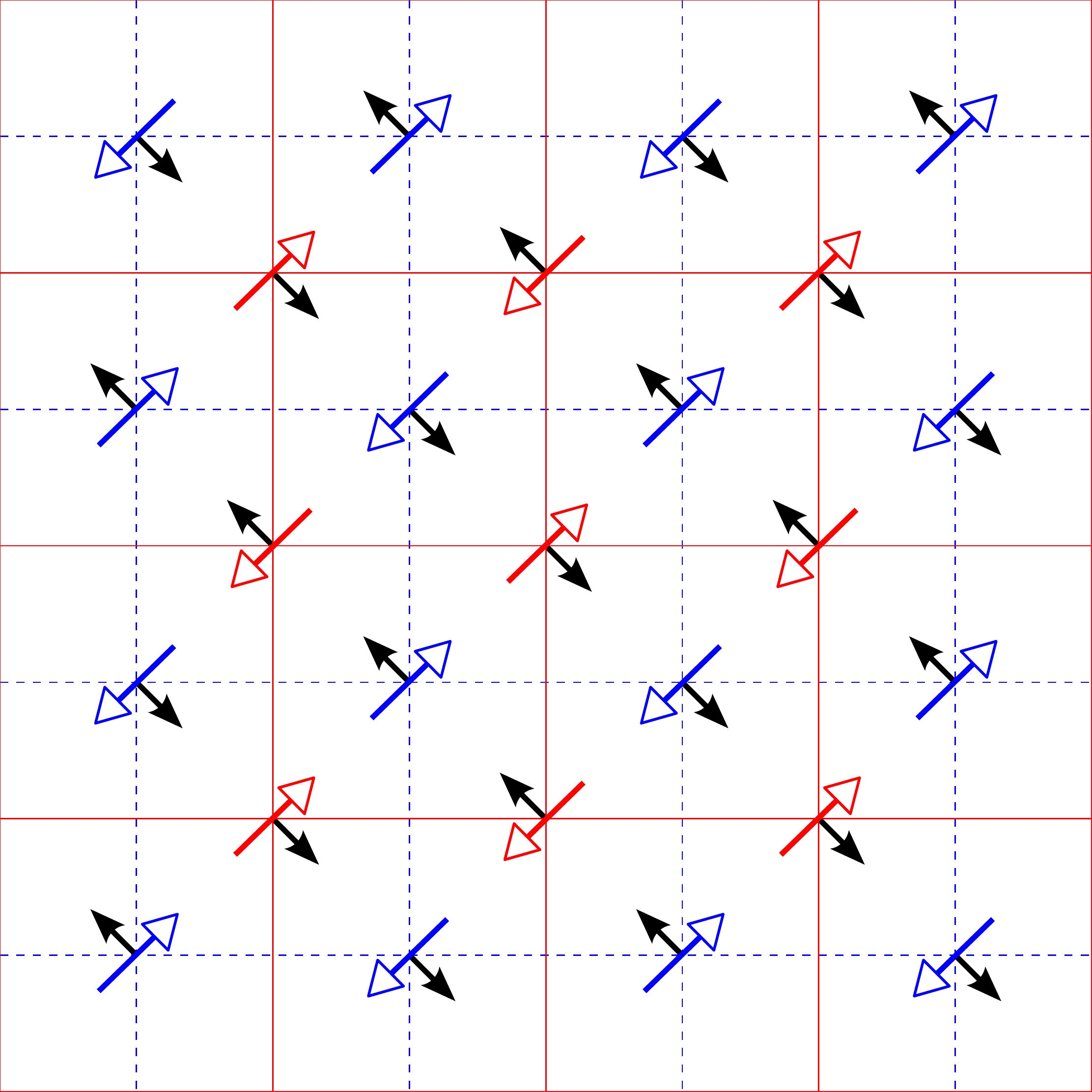}}
    \caption[Planar movement of oxygen ions at AO/AO stacking
      faults]{Planar movement of oxygen ions on opposite sides of
      AO/AO stacking faults: rotation of octahedra in lower layer
      $\mathfrak{A}$
      (\textcolor{red}{$\nearrow\mspace{-8mu}\swarrow$}, outlined
      arrowheads on solid grid) and upper layer $\mathfrak{B}$
      (\textcolor{blue}{$\nearrow\mspace{-8mu}\swarrow$}, outlined
      arrowheads on dashed grid) projected onto interfacial plane and
      resulting movement of oxygens
      ($\boldsymbol{\nwarrow\mspace{-8mu}\searrow}$, filled
      arrowheads) within such plane.  Panels depict (a)
      high\hyp{}energy configuration with movement of neighboring
      oxygen atoms directly toward each other and (b) low\hyp{}energy
      configuration with oxygen atoms moving past each other.}
    \label{fig:stackedOs}
  \end{centering}
\end{figure}

Finally, comparing the results in Table~\ref{tab:formEnergiesIsoSF}
with those in Table~\ref{tab:interEnergiesIsoSF}, we come to the
commonsensical, though not logically necessary, result that the
sequence of stacking faults from the most stable (i.e., lowest
formation energy) to the least stable, directly corresponds to the
sequence of binding strengths, from the strongest attraction to the
weakest.  We also propose a simple physical mechanism for
differentiating between higher and lower energies of formation for the
configurations, potentially reflecting their stability relative to
phase separation into bulk perovskite and bulk A\hyp{}oxide.
Figure~\ref{fig:stackedOs} depicts the planar movement of oxygens in
the AO layers on opposite sides of the stacking fault.  The sense of
rotation of the interfacial octahedra drives constituent oxygen ions
to move in specific patterns within their respective AO layers along
the stacking fault.  For the high\hyp{}energy configurations, $[0^-]$
or $[\pi^+]$, the interfacial oxygen ions from opposite AO layers are
displaced toward each other, despite their natural Coulombic
repulsion, as in Figure~\ref{fig:stackedOs}(a).  For the
low\hyp{}energy configurations in Figure~\ref{fig:stackedOs}(b),
$[0^+]$ or $[\pi^-]$, the same oxygen ions move past each other along
the interface, tending to lessen the approach of like\hyp{}charged
ions toward one another and thus minimizing the energy.  Thus, we find
that the stability of different stacking\hyp{}fault configurations can
be readily understood in terms of simple Coulombic effects.

\subsection{Point defects}
\label{subsec:resultsPoint}

With the energies of the various RP phases now determined, we are
finally in a position to substantiate the claim that, in A\hyp{}rich
stoichiometries of ABO$_3$, the RP phases are preferred over the
formation of point defects.

Interstitials in perovskites are known to be energetically very
unfavorable\cite{balachandran1982otd}.  On the other hand, A\hyp{}rich
stoichiometries can be understood by noting that the addition of an AO
chemical unit is identical to the incorporation of a bulk ABO$_3$
chemical unit followed by the \emph{removal} of a BO$_2$ chemical
unit, the latter step forming a BO$_2$ vacancy complex.  To decide
whether this formation of vacancies is more or less favorable than the
incorporation of excess AO chemical units into planar stacking faults
of the type in RP phases, great care must be taken to accurately
balance stoichiometries.

In the current context, we have been considering the excess AO
chemical units originating from an external reservoir of A\hyp{}oxide.
To maintain a consistent chemical potential, the formation energy of
an isolated BO$_2$ vacancy in bulk perovskite must be computed as the
energy first to remove an extra AO chemical unit from bulk
A\hyp{}oxide, and then to insert it into a large supercell containing
$n$ chemical units of ABO$_3$.  Expressed mathematically, the energy
of formation of this vacancy is
\begin{align} \label{eqn:vAO2}
  E_V &\equiv - E_{\text{AO}} + E_{\text{A}_{n+1}\text{B}_n\text{O}_{3n+1}} - E_{\text{A}_n\text{B}_n\text{O}_{3n}}\\
  &= - E_{\text{AO}} + E_{\text{A}_{m}\text{B}_{m-1}\text{O}_{3m-2}} - E_{\text{A}_{m-1}\text{B}_{m-1}\text{O}_{3(m-1)}}\nonumber\\
  &= \phantom{-} E_{\text{A}_{m}\text{B}_{m-1}\text{O}_{3m-2}} - \left[(m-1) E_{\text{A}\text{B}\text{O}_{3}} + E_{\text{AO}} \right]\nonumber,
\end{align}
where $m \equiv n + 1$.  As the final equality in \eqref{eqn:vAO2}
indicates, the energy of formation of an isolated BO$_2$ vacancy in
bulk perovskite is equivalent to the energy of a BO$_2$ vacancy in an
initial bulk cell containing $m$ chemical units of ABO$_3$
\emph{relative} to the energy of $m - 1$ chemical units of ABO$_3$ and
one chemical unit of AO.

\begin{figure}
  \begin{centering}
    \includegraphics[width=\columnwidth]{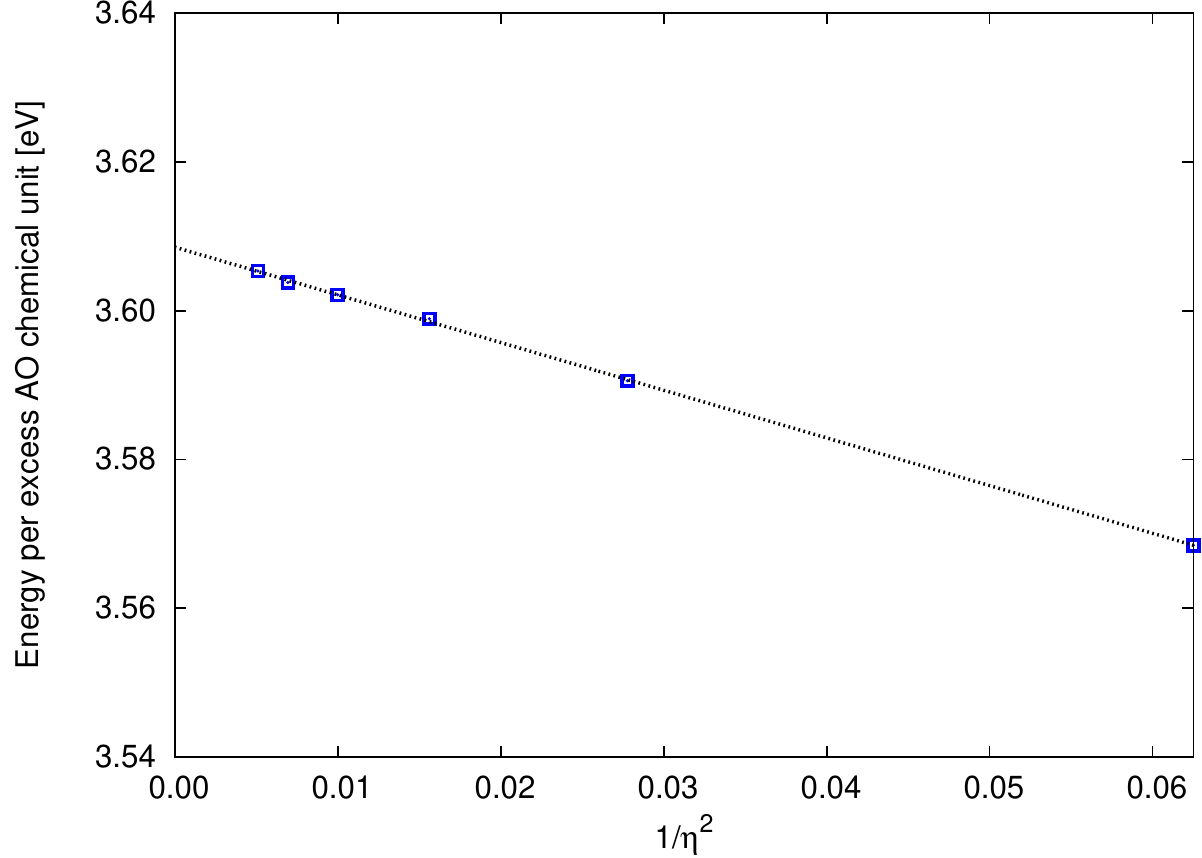}
    \caption[Formation energies of BO$_2$ vacancies]{Formation
      energies of BO$_2$ vacancies as a function of
      inverse\hyp{}square supercell dimension $1 / \eta^2$: extraction
      of energies to infinite supercell (dotted line) to determine
      energy of isolated vacancy.  (See text for definition of
      $\eta$.)}
    \label{fig:pointDefect}
  \end{centering}
\end{figure}

Figure~\ref{fig:pointDefect} presents $E_V$ computed as described
above, for $m = \eta^3$ (corresponding to a series of supercells of
size $\eta = 4,6,\ldots,14$ primitive unit cells on a side, with
between $317$ and $13\,717$ atoms each), plotted as a function of $1 /
\eta^2$.  Following a least\hyp{}squares fit, the unambiguous linear
behavior in the plot indicates that the net interaction decays as $1 /
\eta^2$, with the ordinate intercept ($1 / \eta^2 \rightarrow 0$, or
$\eta \rightarrow \infty$) providing the extracted value of $E_V$ for
the incorporation of excess AO as \emph{fully isolated} BO$_2$ vacancy
complexes.  The result within our model is $E_V = 3.61$~eV per excess
AO unit, far exceeding the energy of formation for even the least
favorable RP phase configuration ($162$~meV per excess AO chemical
unit for $n \rightarrow \infty$ in the $[0^-]$ configuration).  The
positive sign and large magnitude of this energy indicate that
A\hyp{}rich ABO$_3$ will preferentially form planar AO stacking faults
rather than isolated BO$_2$ vacancy complexes.  This result supports
the present work's focus on such stacking faults and RP phases.

\section{Summary and conclusion}
\label{sec:summary}

This work presents a detailed and exhaustive study of the
zero\hyp{}temperature structures of the homologous
A$_{n+1}$B$_n$O$_{3n+1}$ series of Ruddlesden--Popper (RP) phases in a
model antiferrodistortive perovskite for the greatest range of $n$
considered to date.  We also consider, for a specific Glazer system,
\emph{all possible} octahedral orientations within RP phases of the
material.  Our work introduces and follows a general program which may
be replicated for RP phases of perovskites within any Glazer system
and applied in both \emph{ab initio} and empirical potential studies.

We begin our particular considerations with a careful enumeration and
symmetry analysis for all thirty\hyp{}two combinatorially possible
configurations of octahedral rotations within RP phases of an
$a^-a^-a^-$ Glazer system.  Symmetry arguments reduce these
thirty\hyp{}two possibilities to five distinct configurations, which
we denote as $[0^+]$, $[0^-]$, $[\frac{\pi}{2}^+]$, $[\pi^+]$, and
$[\pi^-]$.

We then proceed to determine, within a shell\hyp{}potential model, the
ground\hyp{}state structures of these five symmetry\hyp{}distinct
configurations for all values of $n = 1\ldots30$.  We find that, for
our specific model, only four configurations are actually distinct, as
one of the configurations ($[\frac{\pi}{2}^+]$) is unstable to
immediate relaxation to another configuration ($[\pi^-]$).  After
additional perturbation and quenching tests, we identify three of
these remaining configurations as stable ($[\pi^-]$, $[0^+]$,
$[\pi^+]$), and one as meta\hyp{}stable ($[0^-]$).  In agreement with
experimental observation across a range of perovskites, we find that,
for all values of $n$, all of the configurations of RP phases provide
a much more stable option for the incorporation of excess A\hyp{}oxide
than the formation of point defects, such as BO$_2$ vacancy complexes.
We also find that, within our model, the two most stable
configurations ($[\pi^-]$ and $[0^+]$) lead to stacking faults which
are energetically preferred over phase separation into the bulk
perovskite and bulk A\hyp{}oxide.  Both of these findings are
consistent with the experimental observation of RP phases in a wide
variety of antiferrodistortive perovskites, including titanates,
ruthenates, manganites, and niobates.

In terms of the interactions between stacking faults, we find that the
interaction varies inversely with the distance between excess AO
planes, a result which we believe to be generic, at least for the
$a^-a^-a^-$ system.  For our particular model perovskite, the form of
the interaction between stacking faults is attractive.  In perovskites
with such an attractive interaction, we find that ABO$_3$ with
relatively low stoichiometric excess of species A manifests
intergrowths between the $n = 2$ RP phase and bulk perovskite.  When
such intergrowths are observed experimentally, the interaction is
likely attractive in correspondence with our model material.

Additionally, we observe that the five symmetry\hyp{}distinct
configurations have significant differences in energies of formation:
$\sim$$30$~meV per rotational degree of freedom for our model
perovskite, appreciably greater than the energy scales associated with
orientational disorder in our model perovskite, $4$--$16$~meV\@.  This
leads to an intriguing conjecture.  For low densities of stacking
faults within some perovskites, for a range of temperatures above
their orientational phase transitions, the octahedra in bulk regions
would exhibit randomized behavior about a high\hyp{}symmetry mean
configuration while octahedra near stacking faults would exhibit more
ordered behavior about a lower\hyp{}symmetry mean.  Alternatively, for
higher densities of such faults, this increase in energy scale would
correlate to an increase of the transition temperatures associated
with octahedral rotations.  These effects may be particularly
noticeable in materials with higher orientational energy scales than
our model perovskite, such as lanthanum aluminate with its transition
temperature of $800$~K.

Finally, we propose a simple physical mechanism to explain the strong
dependence of the interfacial energy on the rotational state of the
octahedra: some configurations result in movement of neighboring
like\hyp{}charged oxygen ions directly toward each other and thus are
high in energy, whereas others result in movement of oxygen ions past
each other and thus are low in energy.

\begin{acknowledgments}
  The authors are grateful for interesting conversations with David
  Roundy, Johannes Lischner, and JeeHye Lee.  This work was supported
  by the Cornell Center for Materials Research (CCMR) with funding
  from the Materials Research Science and Engineering Center (MRSEC)
  program of the National Science Foundation (cooperative agreement
  DMR 0520404).
\end{acknowledgments}

\bibliography{paper}

\end{document}